\documentstyle[prd,preprint,aps,eqsecnum,epsfig,%
  amssymb,newlfont]{revtex} 
\newcommand{\mIm}{\,\mathrm{Im}\,}

\newcommand{\be}{\begin{eqnarray}}
\newcommand{\ee}{\end{eqnarray}}
\newcommand{\nee}{\nonumber\end{eqnarray}}

\def\Frac#1#2{\frac{\displaystyle{#1}}{\displaystyle{#2}}} 
\def\lsim{\raise0.3ex\hbox{$\;<$\kern-0.75em\raise-1.1ex\hbox{$\sim\;$}}}
\def\gsim{\raise0.3ex\hbox{$\;>$\kern-0.75em\raise-1.1ex\hbox{$\sim\;$}}}

\begin{document} 
\bibliographystyle{plain} 

%
\preprint{
\begin{tabular}{r}
UAHEP012 \\
UWThPh--2001--13\\
HEPHY--PUB 737 \\ 
IFIC/01-06\\
FTUV-01-03-30\\
\end{tabular}
}
\title{General Flavor Blind MSSM and CP Violation}
\author{\mbox{A. Bartl}$^{1}$, 
 \mbox{T. Gajdosik}$^{2,3}$,
 \mbox{E. Lunghi}$^{4}$,
 \mbox{A. Masiero}$^{5}$,
 \mbox{W. Porod}$^{6}$, 
 \mbox{H. Stremnitzer}$^{1}$,
 \mbox{and O. Vives}$^{6,7}$}
\address{$^{1}$Institut f\"ur Theoretische Physik, 
  Universit\"at Wien, A-1090, Vienna, Austria \\
  $^{2}$ University of Alabama, Tuscaloosa, Alabama 35487, USA\\
  $^{3}$Institut f\"ur Hochenergiephysik der \"Osterreichischen Akademie
  der Wissenschaften\\ A-1050, Vienna, Austria \\
  $^{4}$ Deutches Elektronen Synchrotron DESY, Hamburg\\
  $^{5}$  SISSA -- ISAS and INFN, Sezione di Trieste, I-34013, Trieste, 
Italy\\ 
  $^{6}$ Inst. de F\'{\i}sica Corpuscular (IFIC), CSIC--U. de Val\`encia,
 E-46071, Val\`encia, Spain \\
  $^{7}$ Departament de F\'{\i}sica Te\`orica,
Universitat de Val\`encia, E-46100, Burjassot, Spain}

\maketitle
\begin{abstract}
We study the implications on flavor changing neutral current and $CP$
violating processes in the context of supersymmetric theories without
a new flavor structure (flavor blind supersymmetry).  The low energy
parameters are determined by the running of the soft breaking terms
from the grand unified scale with SUSY phases consistent with the EDM
constraints. We find that the $CP$ asymmetry in $b\to s \gamma$ can
reach large values potentially measurable at B factories, especially
in the low $BR (b \to s \gamma)$ region.  We perform a fit of the
unitarity triangle including all the relevant observables.  In this
case, no sizeable deviations from the SM expectations are found.
Finally we analyze the SUSY contributions to the anomalous magnetic
moment of the muon pointing out its impact on the $b\to s\gamma$ $CP$
asymmetry and on the SUSY spectrum including chargino and stop masses.

\end{abstract}

%
\section{Introduction}
The Standard Model (SM) of electroweak and strong interactions has
been extremely successful in the description of all known high energy
phenomena up to energies of $\cal{O} (100 \ \mbox{GeV})$.  Still, this
impressive theoretical construction is not complete if only because it
does not account for gravitational interactions. Moreover, several
theoretical questions remain unanswered and some cosmological
observations can not be properly accommodated.

From the point of view of theory, the SM includes three independent
gauge coupling constants that in a more complete framework would be
expected to emerge from a single unified parameter.  In turn, this
requirement causes the so--called gauge hierarchy problem: scalar
masses are not protected by any symmetry against radiative corrections
that tend to be of the order of the highest scale present in the
theory.  In a grand unified scenario, this scale is usually close to
the Plank scale; therefore, a great amount of fine tuning is necessary
to keep the Higgs mass close to the electroweak scale.  Furthermore,
in the SM picture, several cosmological observations seem difficult to
account for. In first place, a suitable candidate to reproduce the
required dark matter content of the universe is not provided;
moreover, the tiny $CP$ violation present in the
Cabibbo--Kobayashi--Maskawa (CKM) matrix does not succeed to generate
the necessary baryon--antibaryon asymmetry. Finally in the SM
framework is not possible to include an inflationary stage at the
early universe.

Low energy supersymmetry (SUSY) is the most promising extension of the
SM where all these problems can be successfully solved. Indeed it
stabilizes the gauge hierarchies and successfully unifies the gauge
couplings with remarkably high accuracy.  It includes many possible
dark matter candidates and it can naturally account for inflation as
well. Moreover, the many additional phases present in any SUSY model
can generate the required baryon asymmetry.  Not to mention that any
fundamental theory including gravity almost necessarily contains SUSY
as well (although the scale of SUSY breaking is not forced to be of
the order of the electroweak scale).

For all these reasons, SUSY is one of the most attractive theories
beyond the SM. The so--called Minimal Supersymmetric SM (MSSM) is
obtained adding to the SM spectrum the smallest number of new fields
consistent with SUSY. This recipe does not unambiguously define a
single supersymmetric theory. In fact, to specify completely the
theory, it is necessary to fix the soft breaking terms: this amounts
to 124 parameters at the electroweak scale (luckily enough, most of
this enormous parameter space is already ruled out by phenomenological
constraints).

In this paper, we focus on a certain class of SUSY extensions that we
call {\it flavor blind MSSM}. With this term we refer to a model where
the soft breaking terms at the grand unification (GUT) scale do not
introduce any new flavor structure beyond the usual Yukawa
matrices. These matrices are already present in the superpotential and
are necessary to reproduce correctly the fermion masses and mixing
angles. In this restricted class of models, the number of parameters
is largely reduced and it is therefore possible to perform a complete
phenomenological analysis. Indeed, many features of such models are
shared by most MSSMs. In particular the spectrum and the flavor
conserving processes are not expected to be strongly influenced by
extra flavor structures.

The experimental search for SUSY proceeds through two main lines.  The
main path to establish the existence of low energy SUSY is the direct
search of SUSY particles at present and future colliders with high
enough energy. In addition to these direct searches it is necessary to
perform also the so--called ``indirect searches'' of SUSY particles by
measuring suitable observables with high precision at lower
energies. Virtual SUSY particle contributions appear in the quantum
corrections to characteristic observables and may be traced out if the
experimental and theoretical precisions are sufficient. These indirect
searches of SUSY is particularly important as long as the collider
energies are not, presently, high enough to directly produce the SUSY
particles.

There are two prominent classes of observables which are especially
well suited for probing virtual SUSY particles. These are observables
sensitive to $CP$ violation and observables involving flavor changing
neutral currents (FCNC).  In the context of indirect SUSY searches,
the most interesting $CP$ violating observables are those where the SM
predictions turn out to be very small.  Similarly, the study of FCNC
within SUSY is motivated by the absence of SM tree level
contributions; therefore, one--loop SUSY contributions may be large
enough to give sizeable deviations.

The electric dipole moments (EDMs) of electron and neutron are
well--known examples of flavor conserving observables sensitive to
$CP$ violation.  The SM predictions for these quantities are extremely
small, because the first nonvanishing contributions arise at
higher--loop level. These predictions are several orders of magnitude
below the experimental limits. The SUSY contributions arise already at
one--loop level and can be close to the experimental upper
bounds. Therefore, the electron and neutron EDMs are well suited to
yield important information about SUSY models and can considerably
restrict the allowed parameter regions.  However, in the calculation
of the electron and neutron EDMs it turns out that large cancellations
between the different SUSY contributions can occur. This peculiarity
has to be taken into account when deriving bounds on the SUSY
parameters and phases in specific models. The allowed SUSY parameter
space, especially the phases, can be much larger due to these
cancellations.

An important example of an observable involving FCNC is the branching
ratio of the rare $b$--quark decay $b \rightarrow s \gamma$. The SM
prediction at one--loop level for the branching ratio is comparable in
size with the experimentally measured value. Therefore, a comparison
of the theoretical predictions with the experimental value leads to
considerable restrictions on the parameter space of SUSY models.

In the present paper we study in a systematic way the restrictions on
the SUSY parameters and complex phases which can be derived from the
experimental information on FCNC and on $CP$ violation.  The electron
EDM and the branching ratio of the rare decay $b \rightarrow s \gamma$
constitute the two most severe constraints. We define the SUSY model
at the GUT scale and determine the soft SUSY breaking parameters at
the weak scale by evolving them down with the renormalization group
equations (RGE). We fix $|\mu|^2$ by demanding radiative breaking of
the electroweak $SU(2)_L \times U(1)$ symmetry.  At this scale we
impose the constraints from direct searches and from the
$\rho$--parameter, as well as the requirements of color and electric
charge conservation and the lightest SUSY particle (LSP) to be
neutral.  With these sets of soft SUSY parameters we calculate the EDM
of the electron and the branching ratio of $b \rightarrow s \gamma$
and compare them with the experimental data. The sets in agreement
with the experimental constraints are used to calculate the $CP$
asymmetry of $b \rightarrow s \gamma$ and the SUSY contributions to
the $CP$-violating quantities $\epsilon_K$, $\Delta M_{B_d}$ and
$\Delta M_{B_s}$. With these results we study the modifications of the
so--called unitarity triangle. Finally we calculate the SUSY
contributions to the muon anomalous magnetic moment in order to
quantify the effect of the recent experimental data on the observables
we are interested in.

\section{GUT--inspired MSSM spectrum at $M_W$}


\label{section:spectrum}
In the flavor blind MSSM the most general structure of the soft
breaking parameters at the GUT scale is
\begin{eqnarray}
\label{soft}
& (M_Q^2)_{i j} = M_Q^2\  \delta_{i j}, \ \ \  (M_U^2)_{i j} = M_U^2\
\delta_{i j} , \ \ \
(M_D^2)_{i j} = M_D^2\  \delta_{i j}, & \nonumber \\
&(M_L^2)_{i j} = M_L^2\  \delta_{i j} ,\ \ \  (M_E^2)_{i j} = M_E^2\
\delta_{i j} ,\ \ \ \ \
M_{H_1}^2 ,\ \ \ \ \ \ M_{H_2}^2,\ \ \ &\nonumber \\
& (Y^A_U)_{i j}= A_U e^{i \phi_{A_U}} (Y_U)_{i j}, \ \ \
(Y^A_D)_{i j}= A_D e^{i \phi_{A_D}}
(Y_D)_{i j} ,&\nonumber \\
& (Y^A_E)_{i j}= A_E e^{i \phi_{A_E}} (Y_E)_{i j}, &
\end{eqnarray} 
\noindent
where $i,j$ are family indices, the $Y^A_f$ are trilinear scalar 
couplings and $Y_f$ denote the Yukawa matrices. All the allowed phases are
explicitly written, with the only exception of possible phases in the
Yukawa matrices.  In addition, we have a universal gaugino mass
parameter $M_{1/2}$ that we can take as real while the $\mu$ parameter
in the superpotential is complex. Notice that $M_{H_1}$ and $M_{H_2}$
are only the Higgs soft breaking masses and not the complete Higgs
masses that can be computed from the scalar potential.

As the number of parameters in Eq.~(2.1) is still rather large, for
our present study we will make further simplifying assumptions.  The
first case we consider is the simplest version of the constrained
MSSM, where we take the following independent parameters:
\begin{description}
\item[(I)] $M_{1/2}$, $M_0^2$, $|A_0|$, $\tan\beta$, $\phi_{\mu}$,
$\phi_{A_0}$,
\end{description}
which means that we impose $M_Q^2 = M_U^2 = M_D^2 = M_L^2 = M_E^2 =
M^2_{H_1} = M^2_{H_2} = M_0^2$, $A_U = A_D = A_E = A_0$ and  $\phi_{A_U} =
\phi_{A_D} = \phi_{A_E} =
\phi_{A_0}$. 

The second case refers to the SUSY $SU(5)$ model. In this model the
sfermions are in a $\bar 5$ and a $10$ multiplet and the Higgs
doublets are members of different $5$ multiplets.  We take the
following set of parameters:
\begin{description}
\item[(II)] $M_{1/2}$, $M_{5}^2$, $M_{10}^2$, $M_{H_1}^2$, $M_{H_2}^2$,
$|A_u|$, $|A_d|$, $\tan\beta$, $\phi_{\mu}$,
$\phi_{A_u}$, $\phi_{A_d}$,
\end{description}
where now we have  $M_D^2 = M_L^2 = M_5^2$, $M_Q^2 = M_U^2 = M_E^2
= M_{10}^2$, $A_U = A_u$, $A_D = A_E = A_d$ and  $\phi_{A_U} =
\phi_{A_u}$, $\phi_{A_D} = \phi_{A_E} = \phi_{A_d}$. 

Although the number of parameters in set (II) is significantly larger
than in set (I), the problem can be handled and a full RGE evolution
and an analysis of the low--energy SUSY spectrum is possible.  In our
analysis, we have used two--loop RGEs as given in~\cite{martin:1994zk}
and one--loop masses as given in~\cite{pierce}.

In the following, we are going to discuss the main features of the
low--energy spectrum relevant for $CP$ violating and FCNC observables.
In particular we are interested in electric dipole moments
(EDM), $\varepsilon_K$, $\Delta M_{B_d}$, $\Delta M_{B_s}$, $BR ( b
\to s \gamma)$ and the anomalous magnetic moment of the muon
($a_{\mu^+}$).

The dominant contributions in flavor conserving observables are
mediated by chargino--sneutrino diagrams for the electron EDM and for
$a_{\mu^+}$ and by both chargino--squark and gluino--squark diagrams
for the neutron EDM. Since we are interested in a light SUSY spectrum,
we will also consider sub--dominant neutralino--sfermion contributions
which are important in a part of the parameter space where
cancellations can occur~\cite{cancel,edm:we}.  The main contributions
for flavor changing $CP$ violating observables ($\varepsilon_K$,
$\Delta M_{B_d}$, $\Delta M_{B_s}$, $BR ( b \to s \gamma)$) are given
by up squark--chargino, top--charged Higgs and the usual SM $W$--boson
contributions.  Hence, we are interested in the following part of the
low energy spectrum: $\chi^+$, $\chi^0$, $\tilde g$, $H^+$,
$\tilde{t}$, light $\tilde{q}$ and $\tilde{l}$.

A very good approximation for their masses in terms of the initial
parameters is already given by the solution of the one--loop RGEs
\cite{RGE,bertolini,Kazakov:2000pe}. In tables \ref{tab:ScalarMass1}, 
\ref{tab:ScalarMass3}, \ref{tab:ScalarMass2},
\ref{tab:sfermionfirst},  and \ref{tab:Aparam}, 
we present the numerical solution of these equations for the various
parameters entering the mass formulae for different values of $\tan
\beta$.

A further important parameter entering the mass matrices is $\mu$. We
calculate its modulus from the requirement of electroweak symmetry
breaking, using the complete one--loop corrections for all
particles~\cite{pierce}. To get an understanding of the general
behavior, we use the corresponding tree level formula which reads:
\begin{eqnarray}
 |\mu|^2 &=& \frac{m^2_{H_2} \sin^2 \beta - m^2_{H_1} \cos^2 \beta}
                  {\cos 2 \beta} - \frac{m^2_Z}{2} \ ,
\end{eqnarray}
where we write in lower cases the physical masses at the electroweak
scale. It is well known that the tree level result can be modified by
large one--loop corrections that are taken into account in our
numerical computations.

In first place, we consider the charged Higgs mass which is connected
at tree level with the $\mu$ parameter by:
\begin{eqnarray}
\label{mH+}
m^2_{H^+} &=& m^2_{A^0} + m^2_W = m^2_{H_2} + m^2_{H_1} + 2 |\mu|^2 +
          m^2_W \nonumber \\
      &=& \frac{\tan^2 \beta + 1}{\tan^2 \beta - 1} (m^2_{H_1}- m^2_{H_2}) 
        +m^2_W - m^2_Z
\end{eqnarray}

Eq.~(\ref{mH+}) implies that the ``tree level'' $\tan \beta$
dependence of $m^2_{H^+}$ is always small, and for $\tan \beta \geq 3$
the only important dependence comes through the $m_{H_1}^2$ and
$m_{H_2}^2$ parameters.  Inserting the numbers given
Table~\ref{tab:ScalarMass1} we obtain the result displayed in
Table~\ref{tab:ScalarMass3} for the CMSSM case and in
Table~\ref{tab:ScalarMass2} for $SU(5)$.

Let us first discuss the CMSSM case:
It is clear from Table \ref{tab:ScalarMass3} that the main
contribution stems from the gaugino mass, $M_{1/2}^2$, both for
$|\mu|^2$ and $m_{H^+}^2$. Other relevant contributions come from the
universal scalar mass $M_0^2$ and the $A_0$--$M_{1/2}$ interference
term.  In this framework, the size of the coefficient $c_2$ implies
that within CMSSM $|\mu|$ is, in general, larger than $M_2 \simeq 0.81
M_{1/2}$. The only possible exception to this rule could come from the
negative sign of the $c_3$ coefficient for the $A_0$--$M_{1/2}$
interference. Only with large $\tan \beta$ and a value of $A_0 \geq
5.5 M_{1/2}$ it would be possible to change the above
situation. However, this possibility is ruled out as soon as we
take into account the direct constraints on gaugino and scalar masses
and specially the requirement of the absence of charge and color
breaking minima which forbids large values of $A_0$ \cite{casas}. This
fact has important consequences, in particular it implies that in the
CMSSM the lightest chargino, as well as the lighter two neutralinos,
will be gaugino--like. The behavior of $m_{H^+}^2$ is similar to
$|\mu|^2$, although the contribution from the $M_0^2$ coefficient
$c_1$ is now more important.  Finally, it is obvious from the tables
that with increasing $\tan\beta$, both $|\mu|$ and $m_{H^+}$ decrease.
In the $SU(5)$ scenario we have different scalar masses for the two
Higgs doublets and the particles in different multiplets, as well as
different trilinear terms.
The physical masses at the electroweak scale depend, in this case, on
the values of these parameters at $M_{GUT}$. In Table
\ref{tab:ScalarMass2} we see the dependence of $m_{H^+}$ on these
initial values. We must emphasize that this is only a decomposition of
the coefficients in Table
\ref{tab:ScalarMass3} in different contributions. Therefore, the sum of the
coefficients $c_1$, $c_2$, $c_3$ and $c_4$ in Table
\ref{tab:ScalarMass2} corresponds to the coefficient $c_1$ in Table
\ref{tab:ScalarMass3} and so on. In this decomposition, it is
interesting to notice the negative sign of the coefficient $c_4$. This
means that with a sufficiently large initial value for $M^2_{H_2}$
and a moderate value of $M_{1/2}$ it is possible to have $|\mu|
\leq M_2$ and a large higgsino component in the lightest chargino and
neutralinos; it is so possible to overcome the bounds that forbid this
possibility in the CMSSM case.

\begin{figure}[H]
\begin{center} 
\epsfig{file=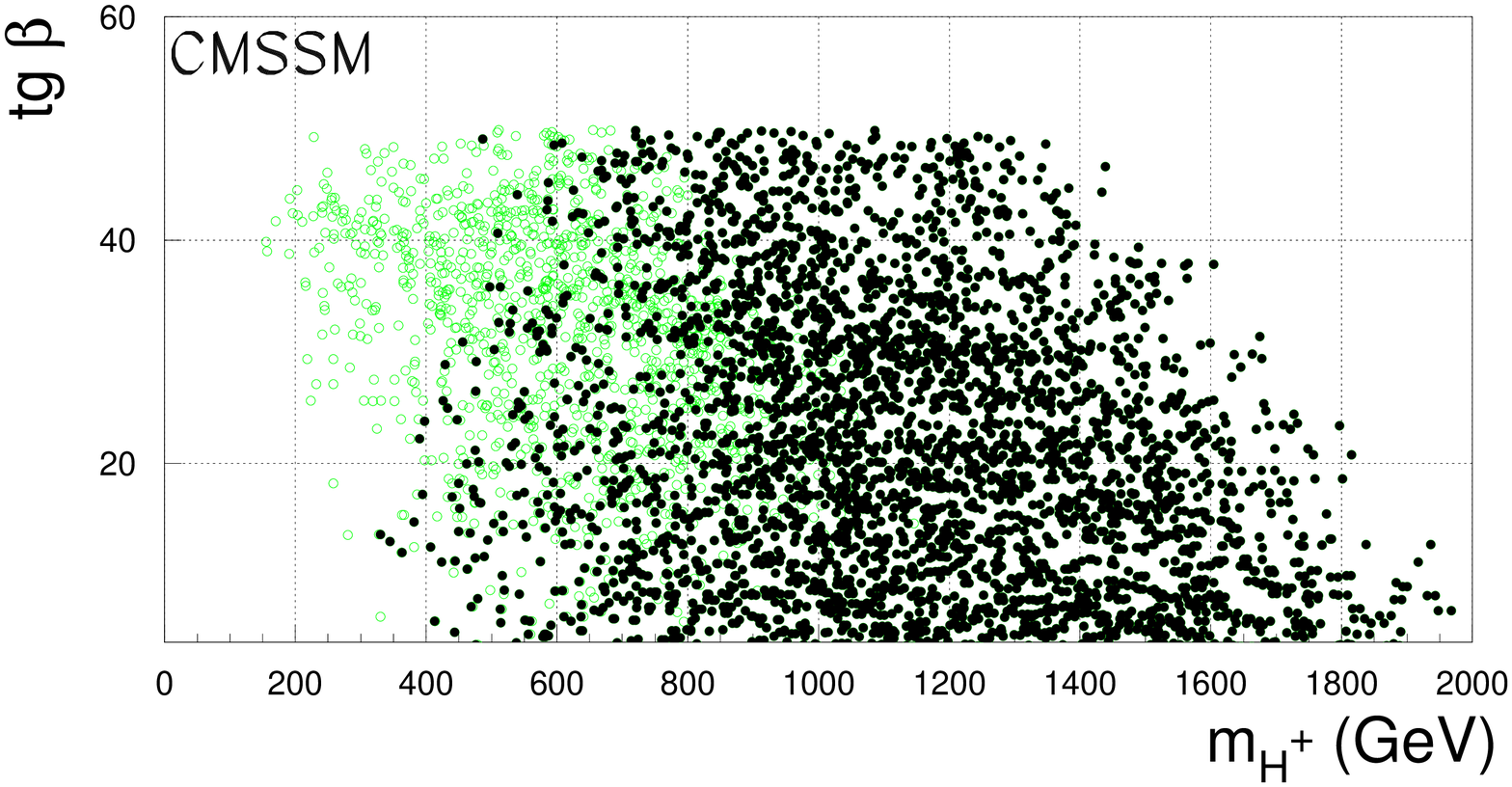,width=0.8 \linewidth}
\epsfig{file=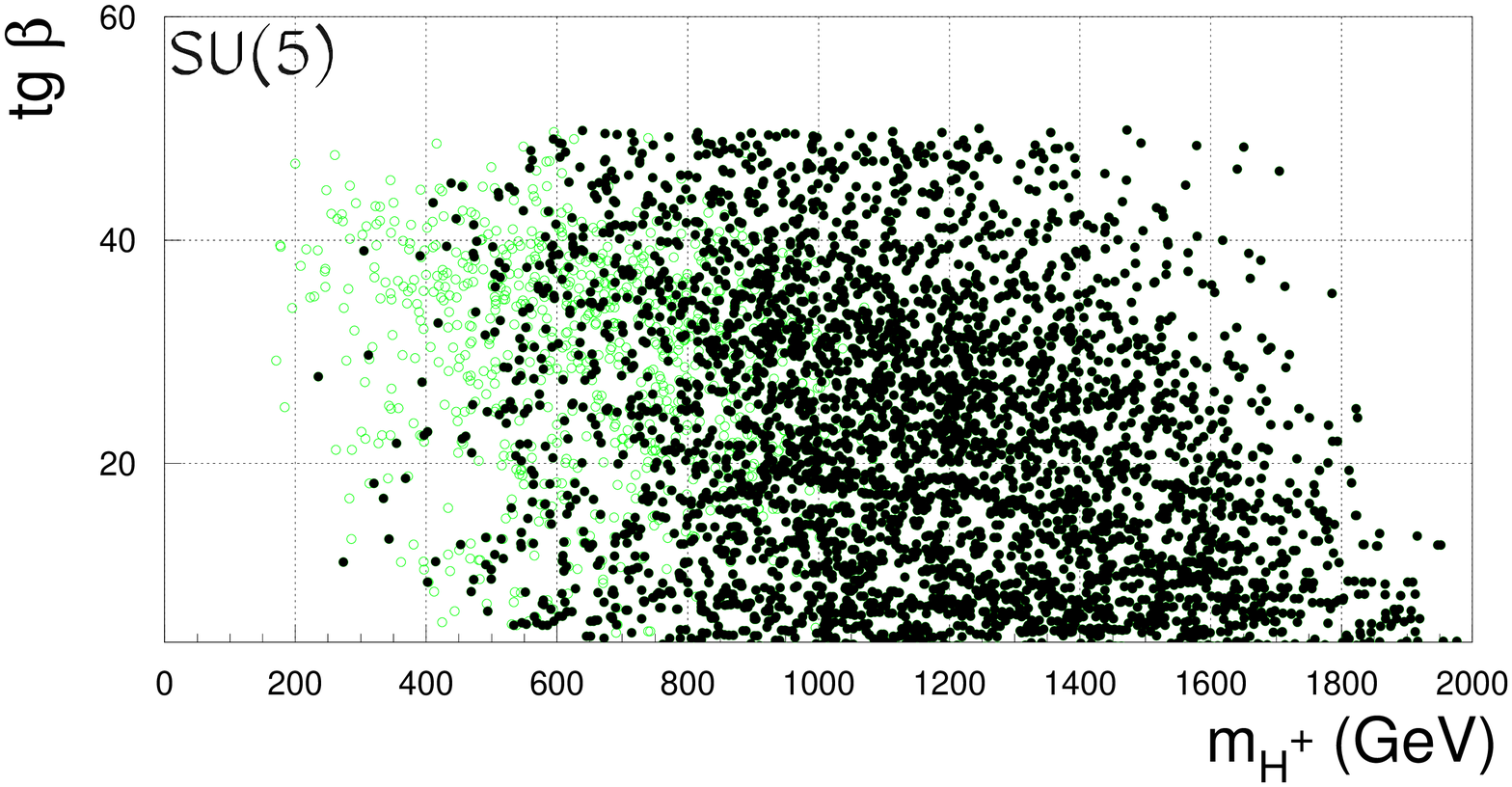,width=0.8 \linewidth}
\caption{Mass of the charged Higgs boson vs $\tan\beta$ in the CMSSM
and in the SU(5)--inspired MSSM. Black dots are points allowed by the $b\to s \gamma$
constraint while open circles fail to satisfy this constraint.}
\label{fig:H+}
\end{center}
\end{figure}

We now present the numerical results obtained using the RGEs at the
two loop level.  In Fig.~\ref{fig:H+} we show the scatter plot of the
mass of the charged Higgs boson versus $\tan \beta$ for the CMSSM and
the SU(5) cases. In this and all the following scatter plots we
vary the scalar and gaugino masses at $M_{GUT}$ in the range $100\ \mbox{GeV} <
M_i < 1\ \mbox{TeV}$,  the trilinear terms 
$0 < |A_d|^2 < M^2_{10} + M^2_5 + M^2_{H_1}$, $0 < |A_u|^2 < 2 \, M^2_{10}
+ M^2_{H_2}$ while their phases are arbitrary.  Finally we take $4 <
\tan \beta < 50$, where the lower bound takes into account the limits
on the lightest neutral Higgs mass.  Moreover, we apply the following
 set of constraints:
\begin{itemize}
\item Absence of charge and color breaking minima and directions
unbounded from below~\cite{casas}.
\item Lower bounds on masses from direct searches \cite{PDG}, in particular
$m_{\chi^+_i} > 90\ \mbox{GeV}$, $m_{\tilde{t}_i} > 90\ \mbox{GeV}$, 
$m_{\chi^0} > 33\ \mbox{GeV}$ and $m_{\tilde{\nu}} > 33\ \mbox{GeV}$.
\item The acceptable range in the $b \to s \gamma$ branching ratio is
between $2 \times 10^{-4}$ and $4.5 \times 10^{-4}$\cite{CLEO}.
\item The lightest supersymmetric particle is neutral.
\item The upper bound on the electron EDM is $|d^e| \leq 4.0 \times 10^{-27} \ e \ \mbox{cm}$.
\end{itemize}
In any $R$-parity conserving MSSM, a further constraint would be the
relic density of the lightest supersymmetric particle, $\Omega_\chi h^2\leq
0.4$.  However, we have not included it because a careful treatment of
this constraint, in the presence of non--vanishing phases, is beyond the
scope of this paper.

In Fig.~\ref{fig:H+}, the black dots fulfill the $b\to s \gamma$
constraint, whereas in the case of the bright open circles this
constraint is violated.  The most relevant feature of these plots is
the heavy mass of the charged Higgs in most of the parameter space,
$m_{H^+} \gsim 400\ \mbox{GeV}$. In the CMSSM case, we see that
$m_{H^+} \geq 400\ \mbox{GeV}$ except for a few exceptions at
intermediate $\tan \beta$.  This is due to the decrease of the $c_2$
coefficient with increasing $\tan \beta$; in this region the $b \to s
\gamma$ constraint is very important too. At small $\tan \beta$, the
chargino contributions cannot usually compete with the charged Higgs
ones and hence the same bound as in two--Higgs doublet models applies,
i.e. $m_{H^+} >250\
\mbox{GeV}$~\cite{borzumati}. With moderate or large values of $\tan \beta$,
the chargino contribution can partially cancel the charged Higgs
contribution and hence lower values are allowed \cite{garisto}. At
larger $\tan \beta$ values, this cancellation is no longer possible
and the $b \to s \gamma$ constraint turns out to be more effective for
low SUSY masses.

In the $SU(5)$ case, a similar situation occurs although more cancellations
are possible because now the charged Higgs and sfermion masses are 
independent at the GUT scale.

\begin{figure}[H]
\begin{center}
\epsfig{file=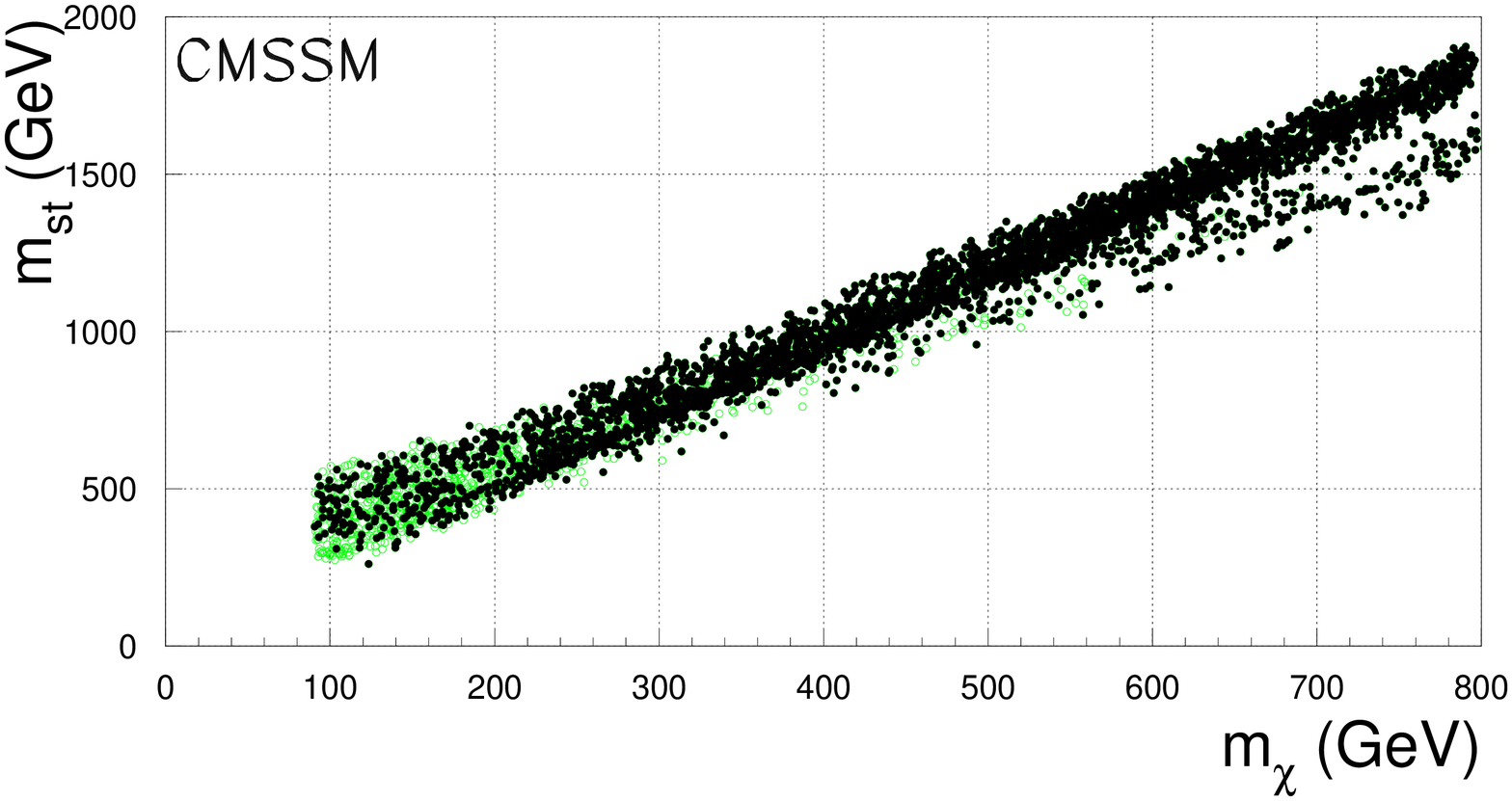,width=0.8 \linewidth}
\epsfig{file=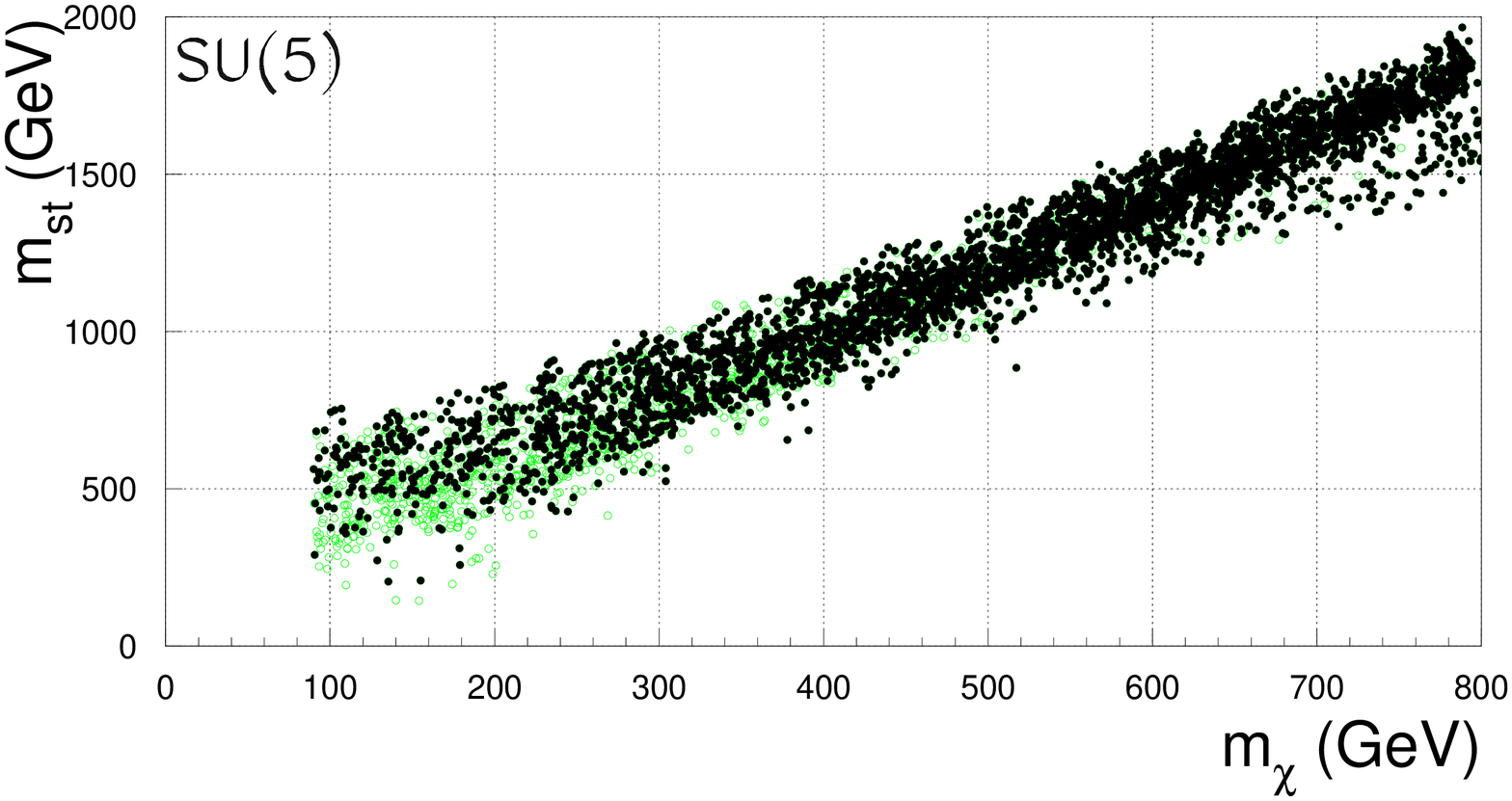,width=0.8 \linewidth}
\caption{Chargino--stop mass correlation in the CMSSM
and in the SU(5)--inspired MSSM. Black dots are points allowed by the $b\to s \gamma$
constraint while open circles fail to satisfy this constraint.}
\label{fig:charstop}
\end{center}
\end{figure}

Beside the charginos and the charged Higgs, that we discussed above,
the stops are particularly interesting in the processes we
consider. Neglecting for the moment the so--called D--terms, one gets
for the masses:
\begin{eqnarray}
m^2_{{\tilde t}_{1,2}} &=&
    \frac{1}{2} \left( m^2_{Q_3} +  m^2_{U_3} + 2 m^2_t 
\mp \sqrt{ ( m^2_{Q_3} -  m^2_{U_3} )^2 + 4 m_t^2 |A_t^* - \mu \cot \beta|^2}
\right).
\label{eq:stopmass1}
\end{eqnarray}
Fig. \ref{fig:charstop} shows a scatter plot of the lightest
chargino versus the lightest stop masses. It is very interesting to
notice the very strong correlation among these masses.  In fact, in
the CMSSM, $|\mu|
\simeq \sqrt{3} M_{1/2}$ is always larger than $M_2 \simeq 0.8 M_{1/2}$
and the lightest chargino, whose mass is bounded by direct searches to
be heavier than $90 \ \mbox{GeV}$, is approximately a gaugino.  On the
other side, we see in Table~\ref{tab:ScalarMass1} that the lightest
stop will always be dominantly right--handed.  Writing
Eq.~(\ref{eq:stopmass1}) as a function of the high scale parameters in
the CMSSM case we find:
\begin{eqnarray}
m^2_{{\tilde t}_{1,2}} &=&
 0.43 M^2_0 + 4.55 M^2_{1/2} + m^2_t + 0.19 M_{1/2} \mbox{Re}(A_0)
 \nonumber \\ &&
 \mp \Frac{1}{2} M_{1/2} \sqrt{2.25 M^2_{1/2} + 1.13 \, M^2_0 + 
20.2 \, m^2_t}
\end{eqnarray}
where a slight dependence on $\tan\beta$ of the coefficients is
neglected.  Using $m_{\chi^+_1} \simeq M_2 \simeq 0.81 M_{1/2}$, we
further get
\begin{eqnarray}
\label{stchar}
m^2_{{\tilde t}_{1,2}} &=&
 0.43 M^2_0 + 6.93 m^2_{\chi^+_1} + m^2_t + 0.23 m_{\chi^+_1} \mbox{Re}(A_0)
 \nonumber \\ &&
 \mp \Frac{1}{2} m_{\chi^+_1} \sqrt{5.23 m^2_{\chi^+_1} + 1.72 \, M^2_0 + 
30.8 \, m^2_t} \ .
\end{eqnarray}

For the mixing the following approximate result can be derived:

\begin{eqnarray}
\label{stopmix}
|\tan 2 \theta_{\tilde t}| \simeq
\frac{2 m_t |0.2 A_0 - (2 - \sqrt{3} \cot \beta  \,\,e^{-i \phi_\mu} )M_{1/2}|}
{|0.37 M^2_0 + 1.3 M^2_{1/2} - 0.13 M_{1/2} \mbox{Re}(A^0)|}
\end{eqnarray}
In case that $A_0 \simeq M_{1/2} \simeq M_0 \simeq m_t$ and
moderate/large $\tan\beta$ one finds that $|\theta_{\tilde t}|\simeq
1.0$.
Therefore, the lighter stop is clearly more ``right handed'' than
the heavier one. Note that for larger $M_{1/2}$ and/or $M_0$ the mixing
angle grows and, thus, the 'right--handed' component of the lighter
top squark increases. An analogue formula can be found for the phase:
\begin{eqnarray}
\tan \phi_{\tilde t} \simeq
\frac{ 0.2 \, \mbox{Im}(A_0) - \sqrt{3} \cot \beta  \sin \phi_\mu M_{1/2}}
     { 0.2 \, \mbox{Re}(A_0)-( 2 - \sqrt{3} \cot \beta  \cos \phi_\mu )M_{1/2}}
\end{eqnarray}
It is obvious from this formula that the $CP$ phase of the top squark
is relatively small at the electroweak scale even if it is maximal at
the GUT scale.  This is a result of the fix--point structure which
governs the corresponding RGEs.

From Eq.~(\ref{stchar}), for $100\ \mbox{GeV} < M_0 < 1\ \mbox{TeV}$ and with 
$m_{\chi^+_1}= 100\ \mbox{GeV}$
we get an allowed range for the stop mass $240\ \mbox{GeV} \lsim m_{\tilde{t}_1} 
\lsim 660\ \mbox{GeV}$. 
As we can see from the plot the correlation between the two masses is
maintained for larger chargino masses.  The ``splitting'' into two
bands is due to the fact that the phase of $\mu$ is rather small and,
thus, it is concentrated around $0$ and $\pi$. One band is more
populated than the other because, according to the EDM constraint,
there is a preferred phase difference between $A_0$ and $\mu$.
 In the case of $SU(5)$, the
main difference is that the Higgs masses are not tied to the
other scalar masses and now may be quite different. This has important
effects in the radiative symmetry breaking; in fact, lower values
of $\mu$ are possible and the lightest chargino can have a
dominant higgsino part.  In scenarios where $|\mu|\lsim M_2$ we
find that the stop masses are somewhat lower compared to the CMSSM
case because of the requirement that $M^2_{H_2} >> M^2_{10}$ and of the
negative sign of the $c_{i4}$ coefficient in
Table~\ref{tab:ScalarMass1} for the $m_{Q_3}$ and $m_{U_3}$
parameters.
In the plot we see that this effect tends to slightly soften the
stop--chargino correlation, however, if we plot $m_{\tilde{t_1}}$
versus $M_2$ much of this correlation is again recovered.  Moreover
the up--type squarks can have a different $A$--parameter value compared to
the down--type squarks and the charged leptons. This leads to a
stronger overlap of the two bands. We must emphasize here that, due to
gluino dominance in the soft--term evolution, this kind of correlation
is general in any RGE evolved MSSM from some GUT initial conditions
assuming that also gaugino masses unify.  In summary, this implies
that the ``light stop and chargino'' scenario
\cite{ToBeFound,ali-london,SUSYNLO,kane1} must be shifted to stop masses in the
range of $250\ \mbox{GeV}$ for chargino masses of $100\
\mbox{GeV}$. As we will see in the next section this has very
important consequences for the searches of low energy FCNC and $CP$
violating effects.

Analogously, a very similar correlation can be obtained for all the
other squarks and Higgs bosons, although it is not as stringent as 
in the stop--chargino case.
We roughly get,
\begin{eqnarray}
m_{\tilde{q}}^2 \simeq 9.3 \times m_{\chi^+_1}^2 + M_0^2.
\end{eqnarray}
It is an interesting fact that the allowed bands are always
wider that in the case of the lighter stop due to the larger coefficient 
of $M_0^2$ and are always above the band plotted in Fig.~2.

\section{Low energy observables}

Indirect searches in $CP$ violation and FCNC 
experiments play a very important role 
in the race for the discovery of SUSY. In these rare processes, SM 
contributions are small and hence supersymmetry is allowed to compete on 
equal ground. We are mainly interested in $CP$ violation experiments, since
new results are coming from
$B$--factories. With this goal, the first observables we must analyze are
the electric dipole moments (EDM) of the quarks and leptons since they  
provide the most stringent constraints on the new supersymmetric phases. 
In this paper, we concentrate on the electron EDM because the constraint it
provides is already very tight and its theoretical computation is independent 
of hadronic uncertainties  that plague the neutron EDM calculation. 
Then, with the information obtained
from the previous analysis of the MSSM spectrum, we address
the study of the main $CP$ violating and FCNC observables consistent
with this EDM constraint. 
In first place, we study the $CP$ violation asymmetry in the $b \to s
\gamma$ decay which, apart from EDMs, is an especially sensitive observable
to the new SUSY phases. Then, we make a full analysis of the unitarity
triangle which includes the study of $\varepsilon_K$, $\Delta M_{B_d}$
and $\Delta M_{B_s}$ as well as the determination of the angles
through the $CP$ asymmetries.  Finally, we complete our discussion
with a study of the anomalous magnetic moment of the muon, whose
measurement has been recently updated at BNL~\cite{brook}.

%
%
\newcommand\quark[2]{{#1}^{}_{#2}}
\newcommand\squark[2]{\tilde{#1}^{}_{#2}}
\newcommand\Charp[1]{\tilde{\chi}^{+}_{#1}}
\newcommand\Charm[1]{\tilde{\chi}^{+c}_{#1}}
\newcommand\Quark[1]{\Psi_{#1}^{}}

\newcommand\squarkq[2]{\tilde{#1}^{*}_{#2}}
\newcommand\Charpq[1]{\bar{\tilde{\chi}}^{+}_{#1}}
\newcommand\Charmq[1]{\bar{\tilde{\chi}}^{+c}_{#1}}
\newcommand\Quarkq[1]{\bar{\Psi}_{#1}^{}}

\newcommand\ckm{\mathrm{K}}
\newcommand\ckmk{\mathrm{K}^{\dagger}}

\newcommand\kc[2]{k^{\tilde{#1}}_{#2}}
\newcommand\lc[2]{l^{\tilde{#1}}_{#2}}

\newcommand\kcq[2]{k^{\tilde{#1} \, *}_{#2}}
\newcommand\lcq[2]{l^{\tilde{#1} \, *}_{#2}}
%
\subsection{EDMs}
\label{sec:EDM}
The EDM of a spin--$\frac{1}{2}$ particle is the coefficient, $d^{f}$,
of the effective operator,
\begin{equation}
{\mathcal L}_{E} = -(i/2) d^{f}
\bar{f} \gamma^{5} \sigma_{\mu\nu} f F^{\mu\nu}
\enspace .
\end{equation}
This $CP$ violating vertex is absent at tree level both in the SM and
in SUSY. It is generated in the SM as a three loop effect
\cite{Czarnecki:1997bu}, whereas SUSY contributions arise generically
as one loop effects. Thus, these contributions are naturally expected
to be much bigger than the stringent limits obtained in various EDM
experiments, namely the neutron~\cite{Altarev:1992cf,Altarev:1996xs},
the Mercury atom \cite{Hg-edm}, and the Thallium atom \cite{eedm}
EDMs, the latter being mostly sensitive to the EDM of the electron.
So far, a fully accepted explanation for the smallness of the EDMs in
SUSY is missing and this fact gives rise to the most severe part of
the so--called supersymmetric $CP$ problem.

It is well--known, since the beginning of the SUSY phenomenology era,
that the effects of $\phi_A$ and $\phi_\mu$ on the electric and
chromoelectric dipole moments of the light quarks imply that
$\phi_{A,\mu}$ should be $<10^{-2}$ \cite{EDMN}, unless one pushes
SUSY masses up to ${\cal{O}}$($1\ \mbox{TeV}$). This strong constraint
led most of the authors dealing with the MSSM afterwards to simply set
$\phi_A$ and $\phi_\mu$ exactly equal to zero. However, in recent
years, the attitude towards the EDM problem in SUSY and the consequent
suppression of the SUSY phases has significantly changed. Indeed,
options have been envisaged allowing for a conveniently suppressed
SUSY contribution to the EDM even in the presence of large (sometimes
maximal) SUSY phases.  Methods of suppressing the EDMs consist in the
cancellation of various SUSY contributions among themselves
\cite{cancel,edm:we}, approximately degenerate heavy sfermions for the first
two generations~\cite{heavy,baek,fully}\footnote{In these models one
has to carefully examine the two--loop contributions with third
generation particles running in the loop~\cite{pilap}.} and
non--universality of the soft breaking parameters at the unification
scale \cite{non-u,EDMfree}.

In our flavor blind scenario, the latter possibility is obviously not
present, and as discussed in the previous section, the SUSY spectrum
is fixed by RGE evolution in terms of the initial conditions at
$M_{GUT}$. We have seen that the first two generations are always
naturally heavier than the third one and hence the second mechanism
can be considered in some regions of the parameter space
\cite{feng}. Still, in a pure MSSM scenario with a SUSY spectrum below
the $\mbox{TeV}$ scale, a cancellation between different contributions
is the main mechanism that can allow large SUSY phases.

In the following, we calculate explicitly the electron EDM and apply the
experimental limits to this value. As shown in \cite{cancel,edm:we},
there exist regions of the parameter space where sizeable phases are
still allowed.  We mainly concentrate on these regions using the
electron EDM to obtain the allowed phases after RGE evolution. We pick
the electron EDM for this procedure because, on the theoretical side,
its calculation is simple and well under control. 
Chargino and neutralino loops are the only SUSY contributions:
\begin{equation}
d^{e} = d^{e}_{\tilde{\chi}^{+}} + d^{e}_{\tilde{\chi}^{0}}
\enspace .
\end{equation}
%
%

The supersymmetric contributions to the EDMs of leptons
and quarks have been calculated in various papers (see 
\cite{cancel,edm:we,EDMN} and references therein).
We use the formulae and numerical results as given in~\cite{edm:we}.
%
%
%
%
The chargino contribution to the electron EDM can be brought to the
very simple form:
\begin{eqnarray}
\frac{1}{e} d^{e}_{\tilde{\chi}^{+}}
&=&
\frac{\alpha_{em} m_{e} \tan\beta \mIm[M_2 \mu]}
     {4 \pi \sin^{2}\theta_{W}(m_{\tilde{\chi}^{+}_{2}}^2 -
m_{\tilde{\chi}^{+}_{1}}^2 ) }
\times \frac{1}{m_{\tilde{\nu}_e}^{2}}
\left( F_3 (r_1) - F_3 (r_2) \right)
\label{EDMchargino}
\end{eqnarray}
where $r_{i} = m_{\tilde{\chi}^{+}_{i}}^2 / m_{\tilde{\nu}^{}_{e}}^2$ and
the loop function $F_3 ( r_i)$ can be found in the appendix.

The neutralino contribution to the electron EDM involves complex
neutralino and sfermion mixings. Hence, its expression is not
especially simple~\cite{edm:we}:
\begin{equation}
\frac{1}{e}
d^{f}_{\tilde{\chi}^{0}}
=
 -\frac{Q_{f}}{8\pi}
  \frac{\alpha}{\sin^{2}\theta_{W}}
\sum_{k=1}^{4} \sum_{m=1}^{2} \eta^{f}_{mk} \times
  \frac{m_{\tilde{\chi}^{0}_{k}}}{m^{2}_{\tilde{f}_{m}}}
  F_4(
\frac{m^{2}_{\tilde{\chi}^{0}_{k}}}{m^{2}_{\tilde{f}_{m}}} )
\label{EDMneutralino}
\end{equation}
where
\begin{eqnarray}
\label{eta}
\eta^{f}_{mk}
&=&
(-1)^{m} \sin 2\theta_{\tilde{f}}
  \mIm[ ( (h^{f}_{Lk})^{2} - f^{f}_{Lk} f^{f*}_{Rk} )
        e^{-i \varphi_{\tilde{f}}}]
\nonumber\\ &&
- ( 1 - (-1)^{m} \cos2\theta_{\tilde{f}})
  \mIm[ h^{f}_{Lk} f^{f*}_{Lk} ]
\nonumber\\ &&
- ( 1 + (-1)^{m} \cos2\theta_{\tilde{f}})
  \mIm[ h^{f}_{Lk} f^{f}_{Rk} ]
\enspace ,
\end{eqnarray}
in terms of the neutralino--sfermion couplings,
\begin{mathletters}
\begin{eqnarray}
h^{e}_{Lj}
&=&
- Y_{e} ( \cos\beta N_{3j} + \sin\beta N_{4j} )
\enspace ,
\label{couplings:hsd}
\\
h^{e}_{Rj}
&=&
- Y_{e} ( \cos\beta N^{*}_{3j} + \sin\beta N^{*}_{4j} )
= h^{e \, *}_{Lj},
\\
f^{f}_{Lj}
&=&
- \left[
    Q_{f} \sin 2\theta_{W} N^{*}_{1j}
  + ( 1 - 2 Q_{f} \sin^{2}\theta_{W} ) N^{*}_{2j}
  \right]
  / ( \sqrt{2} \cos\theta_{W} )
\enspace ,
\\
f^{f}_{Rj}
&=&
  \left[
    Q_{f} \sin 2\theta_{W} N_{1j}
  + ( - 2 Q_{f} \sin^{2}\theta_{W} ) N_{2j}
  \right]
  / ( \sqrt{2} \cos\theta_{W} )
\enspace .
\end{eqnarray}
\end{mathletters}%
with $Y_{e} = m_{e}/(\sqrt{2} \, m_{W} \cos\beta)$ and
the definitions for the neutralino mixing matrix $N_{\alpha j}$ and the
selectron mixing angle $\theta_{\tilde{f}}$ and phase $\varphi_{\tilde{f}}$
as well as the loop function $F_4$ can be found in the Appendices.

With the help of Eqs.~(\ref{EDMchargino}) and (\ref{EDMneutralino}) we can
readily calculate the electron EDM. If both contributions are separately
required to satisfy the experimental bound, $(1.8 \pm 1.2 \pm 1.0) \times
10^{-27} \mbox{$e$ cm}$, we obtain, from the electron EDM constraint,
$\phi_{\mu} \lsim 0.01$ and $\phi_A \lsim 0.3$
for sfermion and gaugino masses of $100\ \mbox{GeV}$ at $M_{GUT}$. 
Notice that the constraint on $\phi_A$ is always less stringent than the 
one on $\phi_\mu$ because the latter contributes also through chargino
mixing and it is enhanced by a $\tan \beta$ factor in the down squarks
and charged sleptons mass matrices.

\begin{figure}[H]
\begin{center}
\epsfig{file=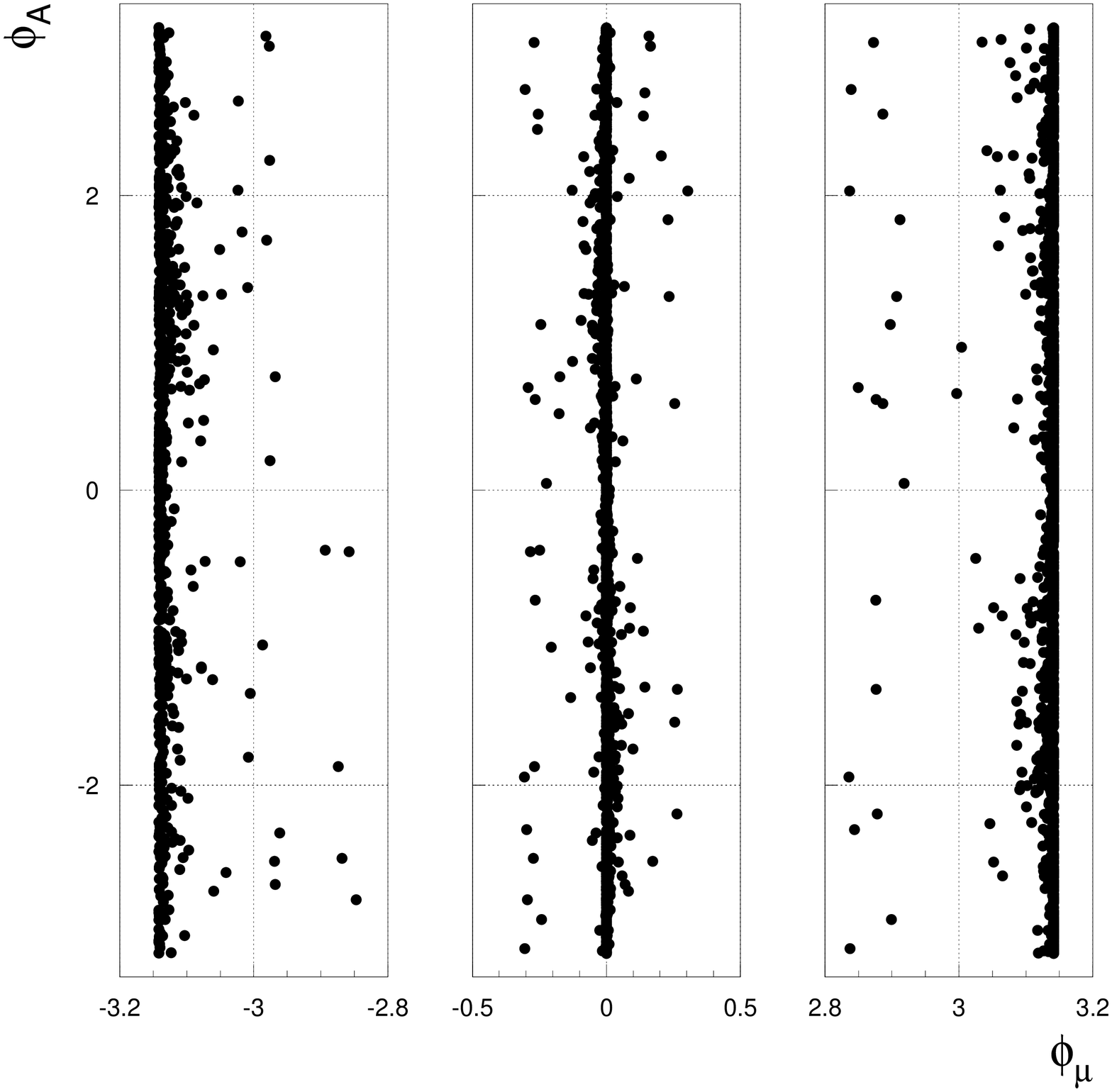,width=0.6 \linewidth}
\epsfig{file=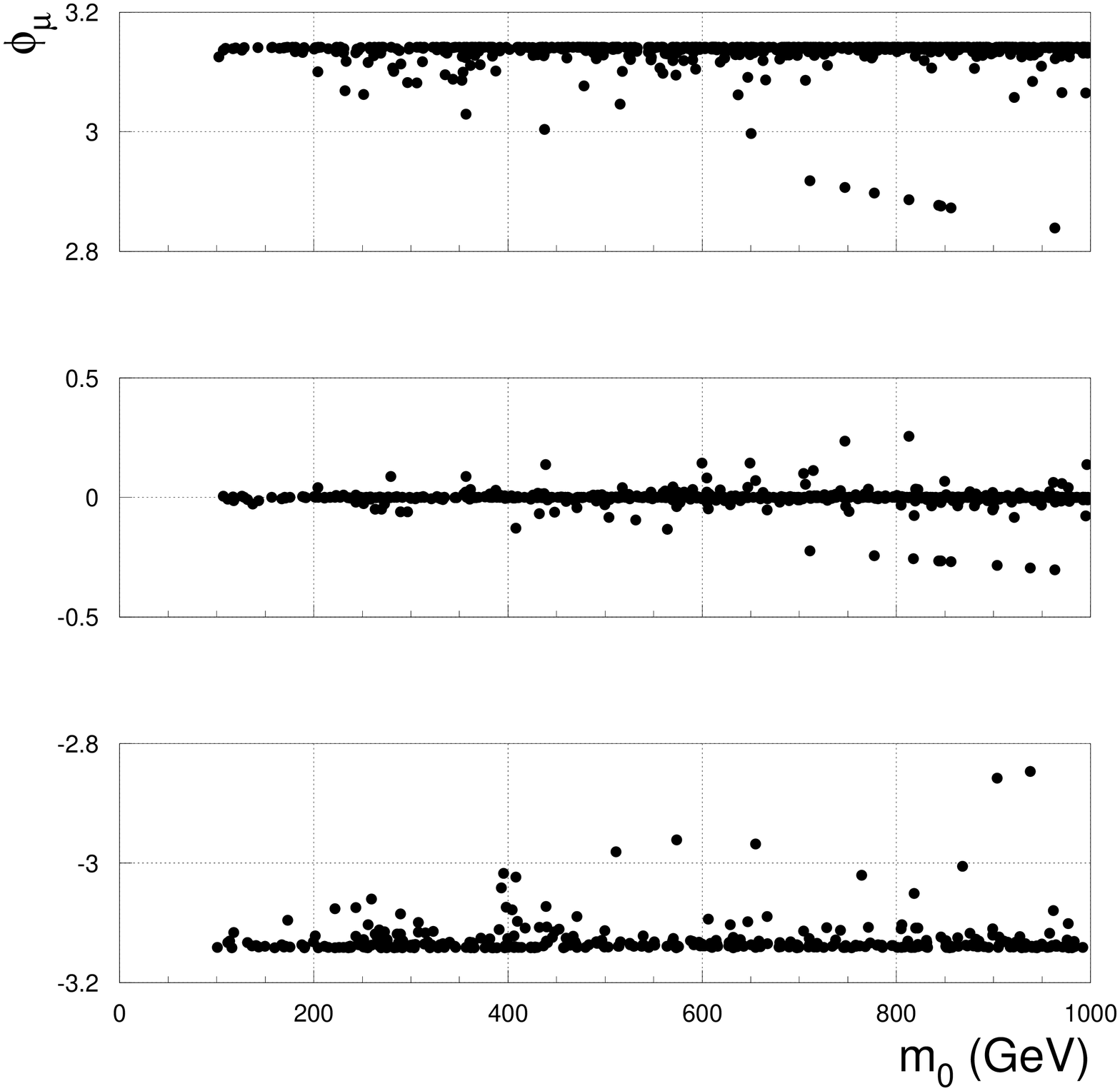,width=0.6 \linewidth}
\caption{Values of the SUSY phases allowed by the EDM constraints, and 
$\phi_\mu$--$M_0$ correlation.}
\label{fig:phases-edm}
\end{center}
\end{figure}

However, if both
contributions are considered simultaneously,  a negative interference
occurs in particular regions of the parameter space and larger phases are
allowed under these special conditions.
In Fig.~\ref{fig:phases-edm} we show the allowed regions for $\phi_\mu$ and
$\phi_{A_e}$ as specified at $M_{GUT}$, and the correlation of $\phi_\mu$
with the scalar mass, $M_0$. As we can see in these
figures,
it is possible to find any value for $\phi_{A_e}$ although there is a
correlation with the value of $\phi_\mu$.
The value of $\phi_\mu$ itself is
much more constrained; however, values up to $\phi_\mu=0.4$ are still allowed,
especially for regions of relatively large masses.
This may appear surprising, given that the usual values quoted from EDM
cancellations are $\phi_\mu \lsim 0.1$ \cite{cancel,edm:we}.
However we must take into account
that in our analysis we do not fix a priori the supersymmetric scale, but
scan the whole range of parameters and so these large phases correspond to
relatively large sfermion masses, up to $1\ \mbox{TeV}$.

To finish our discussion, we add a short comment on neutron and
mercury atom EDMs that should also be included in a complete analysis.
Unfortunately, the prediction for the EDM of the neutron depends on
the specific description of the neutron as a quark bound state. The
first estimates were based on the non--relativistic SU(6) quark model
with $d^{}_{N} = (4/3) d^{d}_{} - (1/3) d^{u}_{}$, where the other QCD
contributions to $d^{}_{N}$ were estimated by a Naive Dimensional
Analysis \cite{Manohar:1984md}.  Another estimate, based on the
measurements of the spin structure of the proton, were made in
\cite{Ellis:1996dg}.  Both estimates give different results for
$d^{}_{N}$ as has been shown in \cite{edm:we}.  A third
approach~\cite{Pospelov:2001bw} uses QCD sum rules.  Another
measurement of EDMs regards the atomic EDM of
${}^{199}\mathrm{Hg}$. Ref.~\cite{olive} uses this result to restrict
the MSSM parameter space.  It is not clear whether it may be possible
to find parameter regions where all the EDM constraints are
simultaneously satisfied~\cite{olive,Abel:2001mc}. An analysis of the
EDMs of electron, neutron and ${}^{199}\mathrm{Hg}$ with implications
for measuring the phases at an $e^+e^-$ linear collider are given
in~\cite{barger} (concerning chargino searches at LEP in the presence
of complex MSSM parameters see \cite{rosiek}). On the other hand, for
our present analysis we can restrict ourselves to the inclusion of
only the electron EDM, hence providing conservative bounds on the
allowed SUSY parameter space. The further inclusion of the neutron and
${}^{199}\mathrm{Hg}$ EDMs is beyond the scope of the present
analysis.

In summary, we have seen in this section that EDM constraints allow for 
sizeable SUSY phases in some regions of the parameter space where a negative
interference takes place. Even in these special regions, $\phi_\mu$
is tightly constrained, while $\phi_A$ is essentially unconstrained. 
In the next sections, we will analyze the effects
of these restricted phases in different $CP$ violation observables.



\subsection{$b \to s \gamma$}
The inclusive radiative decay $B \rightarrow X_s \gamma$ is an extremely
useful tool in testing the FCNC structure of the SM and its possible
extensions. At $95\%$ C.L., its total decay width is restricted to lie
inside the experimentally allowed region~\cite{CLEO},
\begin{equation}
2 \times 10^{-4} \leq \hbox{BR} (B \rightarrow X_s \gamma)
                 \leq 4.5 \times 10^{-4},
\label{bsglimit}
\end{equation}
which is expected to be further reduced within the next few months.
The above constraint turned out to be a great challenge for the
general MSSM parameter space because of the presence of flavor
changing couplings otherwise unconstrained. Even in the more
predictive class of models considered in this paper, it is easy to
exceed the bound (\ref{bsglimit}) with large $\tan \beta$ and a light
superpartner spectrum. In particular, a key role is played by the
charged Higgs, lightest stop and lightest chargino loops. These pieces
are proportional to the CKM mixing matrix and experimentally the
masses of the particles involved are just constrained to be heavier
than approximatively $100\ \mbox{GeV}$.  In general, they provide the
bulk of the SUSY contributions to this decay. On the other hand, to
have sizeable gluino contributions new flavor structures other than
the CKM matrix are required (see for instance Ref.~\cite{bertolini}).
Indeed, in more general SUSY models, gluino contributions are very
important and can be even dominant~\cite{gabbiani}. However, in our
flavor blind scenario, they are typically subdominant. In a similar
way, diagrams involving neutralino exchange can be safely ignored. For
these reasons, we will focus on chargino and charged Higgs exchange in
the remaining of this section.

A first important issue concerns the relative sign of the $W$--top loop with
respect to the stop--chargino and $H^+$--top contributions. Notice that the
latter contribution has always the same sign as the SM
while the former can interfere constructively or destructively
in such a way that, in case of strong cancellations,
the allowed chargino and charged Higgs masses can be very
close to the direct search lower bounds.
In the large $\tan \beta$ region the relative sign of the
chargino mediated diagram is given by $-\hbox{sign}(A_t \mu)$.
Since the value of $A_t$ at the scale $m_W$ in the MSSM with
RGE running from $M_{GUT}$ to the electroweak scale
is essentially determined by $- M_{1/2}$ (see table \ref{tab:Aparam}),
it is clear that $\mu>0$ implies destructive interference of the chargino
contribution.

A complete NLO analysis is available only for the SM~\cite{SMNLO} and
for the two--Higgs--doublet--model~\cite{2HDMNLO} while only partial
results are available for the
MSSM~\cite{SUSYNLO,greub,SUSY1NLO,carena}.  The correct approach would
be to properly take into account the complete SUSY contributions both
to the LO and NLO matching conditions. Given that the NLO results in
SUSY are provided only under particular assumptions on the SUSY mass
spectrum, we prefer to include only the LO matching conditions and to
perform a high statistics scanning of the parameters at the GUT scale.
Our choice is also supported by the analysis presented in
Ref.~\cite{deboer}, where it was pointed out that one of the main
effects of the improved NLO computation of
Refs.~\cite{SUSYNLO,SUSY1NLO,carena} is to reduce the scale
uncertainties while the central values of the predicted BR does not
undergo dramatic changes.  Moreover the other observable we are
interested in, namely the $CP$ asymmetry, is predicted to be
negligibly small in the SM and it is still far to be experimentally
detected.  For these reasons we prefer to use the NLO SM analysis and
to include SUSY effects via their contributions to the LO matching
conditions.

The effective Hamiltonian which describes the transition
$b\rightarrow s \gamma$ in the SM is given by
\begin{equation}
H_{eff}^{b\rightarrow s \gamma} =
- {4 G_F\over \sqrt{2}} K_{tb} K_{ts}^* \sum_{i=1}^8 C_i\cdot Q_i 
\label{bsgEH}
\end{equation}
where the operator basis is defined as follows:
\begin{eqnarray}
Q_1&=& (\bar{s}_{L\alpha} \gamma_\mu c_{L\beta})
(\bar{c}_{L\beta} \gamma^\mu b_{L \alpha}),\nonumber\\
Q_2 &=& (\bar{s}_{L \alpha} \gamma_\mu c_{ L\alpha})
(\bar{c}_{ L\beta} \gamma^\mu b_{ L\beta}),\nonumber\\
Q_3 &=& (\bar{s}_{ L\alpha} \gamma_\mu b_{L \alpha})\sum_{q=u,d,s,c,b}
(\bar{q}_{L \beta} \gamma^\mu q_{L \beta}),\nonumber\\
Q_4&=& (\bar{s}_{L \alpha} \gamma_\mu b_{L \beta})
\sum_{q=u,d,s,c,b}(\bar{q}_{L \beta} \gamma^\mu q_{L \alpha}),\nonumber\\
Q_5&=& (\bar{s}_{L \alpha} \gamma_\mu b_{L \alpha})
\sum_{q=u,d,s,c,b}(\bar{q}_{L \beta} \gamma^\mu q_{R \beta}),\nonumber\\
Q_6&=& (\bar{s}_{L \alpha} \gamma_\mu b_{ L\beta})
\sum_{q=u,d,s,c,b}(\bar{q}_{R \beta} \gamma^\mu q_{R \alpha}),\nonumber\\
Q_7&=&\frac{e m_b}{16 \pi^2}\bar{s}_{L\alpha} \sigma_{\mu \nu}
      b_{R\alpha} F^{\mu \nu},\nonumber\\
Q_8& =& \frac{g_s m_b}{16 \pi^2}\bar{s}_{L\alpha}
T_{\alpha \beta}^a \sigma_{\mu \nu} b_{R \beta} G^{a \mu \nu},
\label{bsgOB}
\end{eqnarray}
where $\alpha$, $\beta$ are color indices and $a$ labels the $SU(3)$
generators. In any MSSM the above basis must be extended to include
\begin{itemize}
\item the opposite chirality operators, obtained interchanging left and right
      fields,
\item the scalar and pseudoscalar operators,
      in which no $\gamma_\mu$ structure is present,
\item the tensor operators, characterized by the presence of the
      $\sigma_{\mu\nu}$ tensor.
\end{itemize}
However the SUSY contributions to the Wilson coefficients (WCs)
of this extended operator
basis turn out to be exceedingly small in our framework due to the
lack of new flavor changing structure other than the CKM matrix.
Moreover the WCs of the operators $Q_{1,...,6}$ are not sizably
modified in any R-parity conserving SUSY theory.  For these reasons we
have only to deal with the SUSY contributions to the operators $Q_7$
and $Q_8$.  The values of the LO WC's at the $m_W$ scale are
\begin{eqnarray}
C_{7,8} (m_W) &=& C_{7,8}^W (m_W) + C_{7,8}^{H^+} (m_W) +
C_{7,8}^{\chi} (m_W), \\ C_7^W (m_W) &=& -{3\over 2} x_t \left[\frac{2}{3}
F_1(x_t)+ F_2(x_t) \right], \\ C_8^W (m_W) &=& -{3\over 2} x_t
F_1(x_t), \\ C_7^{H^+} (m_W) &=& -{1\over 2} {x_t\over x_H}
\left\{{1\over \tan^2 \beta} \left[\frac{2}{3} F_1({x_t\over x_H})+
F_2({x_t\over x_H}) \right] + \left[\frac{2}{3} F_2({x_t\over x_H})+
F_1({x_t\over x_H}) \right]\right\},\\ C_8^{H^+} (m_W) &=& -{1\over 2}
{x_t\over x_H} \left\{{1\over \tan^2 \beta} F_1({x_t\over
x_H})+F_2({x_t\over x_H})\right\},\\ C_7^\chi (m_W) &=&
\sum_{\alpha,\alpha'}\sum_{i=1}^2 \sum_{a=1}^6 {K_{\alpha b}
K^*_{\alpha' s} \over K_{tb} K_{ts}^*} \left\{{1\over s_i}
G^{(\alpha',a)i*} \left[ \frac{2}{3} F_1({z_a\over s_i})+ F_2({z_a\over s_i})
\right] G^{(a,\alpha)i} - \right. \nonumber \\ && \left. {1\over z_a}
{1\over \sqrt{2} \cos \beta} G^{(\alpha',a)i*} \left[ \frac{2}{3}
F_4({z_a\over s_i})+ F_3({z_a\over s_i}) \right] H^{(a,\alpha)i}
\right\}, \\ C_8^\chi (m_W) &=& \sum_{\alpha,\alpha'}\sum_{i=1}^2
\sum_{a=1}^6 {K_{\alpha b} K^*_{\alpha' s} \over K_{tb} K_{ts}^*}
\left\{{1\over s_i} G^{(\alpha',a)i*} F_1({z_a\over s_i})
G^{(a,\alpha)i} - \right. \nonumber \\ && \left. {1\over z_a} {1\over
\sqrt{2} \cos \beta} G^{(\alpha',a)i*} F_4({z_a\over s_i})
H^{(a,\alpha)i} \right\},
\label{SUSYWC}
\end{eqnarray}
where $K_{\alpha q} G^{(\alpha,k)i}$ represents the coupling of the
chargino $i$ and the squark $k$ to the left--handed down quark $q$
and $m_q/(\sqrt{2} m_W \cos \beta) K_{\alpha q} H^{(\alpha,k)i}$ the
coupling of the chargino $i$ and of the squark $k$ to the
right--handed down quark $q$. 
These couplings, in terms of the standard mixing matrices defined in
the appendices are~\cite{haber}
\begin{eqnarray}
G^{(\alpha, k) i} &=& \left( \Gamma_{U L}^{k \alpha} V_{i 1}^{*} -
\frac{m_{\alpha}}{\sqrt{2} m_W \sin \beta} \Gamma_{U R}^{k \alpha}
V_{i 2}^{*}\right)\nonumber \\
H^{(\alpha,k)i} &=& - U_{i 2} \Gamma_{U L}^{k \alpha}.
\end{eqnarray}
Moreover we have $x_q= m_q^2/m_W^2$, $x_H = m_{H^+}^2/m_W^2$, $z_k =
m_{\tilde{u}_k}^2/m_W^2$ and $s_i =m_{\tilde{\chi}_i}^2/m_W^2$.  The
explicit expressions for the loop functions can be found in the
appendices.

Following the analysis presented in Ref.~\cite{Kagan} we consider the ratios
\begin{equation}
\xi_{7,8} \equiv {C_{7,8} (m_W) \over C_{7,8}^W (m_W)}
\end{equation}
and we write the following numerical expression for the
$B\rightarrow X_s \gamma$ branching ratio
\begin{equation}
{BR(B\rightarrow X_s \gamma)\over BR(B\rightarrow X_c e \nu)} =
 1.258+ 0.382 |\xi_7|^2 + 0.015 |\xi_8|^2 + 1.395 \hbox{Re} \xi_7 +
0.161 \hbox{Re} \xi_8 + 0.083 \hbox{Re} \xi_7 \xi_8^*.
\label{bsgBR}
\end{equation}
Eq.~(\ref{bsgBR}) is computed taking into account the SM NLO matching
conditions, fixing the scale of the decay to $m_b$ and
imposing the condition that the photon energy be above the threshold
$E_\gamma \geq (1-\delta) m_b/2$ with $\delta=0.9$ (see
Ref.~\cite{Kagan} for further details).

An especially interesting observable is the $CP$ asymmetry in the
partial width,
\begin{equation}
A_{CP}^{b\rightarrow s\gamma} =
{BR(\bar B\rightarrow X_s \gamma) - BR(B\rightarrow X_s \gamma) \over
 BR(\bar B\rightarrow X_s \gamma) + BR(B\rightarrow X_s \gamma)}.
\end{equation}
This asymmetry is predicted to be exceedingly small in the
SM~\cite{soares} and therefore is sensitive to the presence of new
sources of $CP$ violation.  In particular, the new SUSY phases,
$\phi_\mu$ and $\phi_A$, are associated with chirality changing
operators. Hence, we can expect large effects in chirality changing
decays, as EDMs or $b \to s \gamma$ while their effects are screened
in processes which are dominantly chirality conserving
\cite{flavor,CPcons}. 
Therefore, this is one of the best observables, apart from
EDMs, to find effects of non--vanishing SUSY phases
\cite{baek,hou,aoki-cho,goto}.

According to the analysis of Ref.~\cite{kagan1} we write the following
numerical expression for the $CP$ asymmetry
\begin{equation}
A_{CP}^{b\rightarrow s\gamma} = {\hbox{Im} \left[ 1.06 \; C_2 C_7^*
-9.52 \; C_8 C_7^* +0.16 \; C_2 C_8^* \right] \over |C_7|^2}
\end{equation}
where the WCs are all evaluated at the scale $m_b$.

\begin{figure}[H]
\begin{center}
\epsfig{file=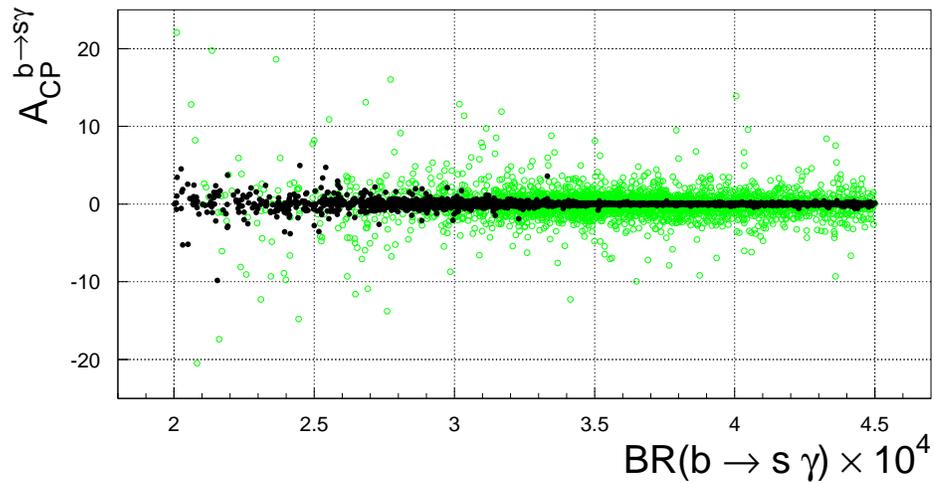,width=0.8 \linewidth}
\caption{$CP$ asymmetry vs total width of the decay $b \rightarrow s \gamma$.
The empty circles are computed without any restriction on the phases.
The filled black ones show the impact of the EDM's constraints.}
\label{fig:aCP}
\end{center}
\end{figure}

The results of the analysis of the $CP$ asymmetry in our scenario are
shown in Fig.~\ref{fig:aCP}. In this figure, open circles represent
points of the parameter space with no restriction on SUSY phases,
while black dots are points satisfying the electron EDM constraint as
explained in the previous section. As expected EDM constraints have a
strong impact on the asymmetry.  Without any restriction on the
phases, it is possible to achieve asymmetries between $5\%$ and $10\%$
for any value of the branching ratio; the much higher allowed values
($10\% \div 20\%$) are due to the smallness of the branching ratio and
not to the underlying structure of the theory.  Once the EDM's bounds
are imposed, the average value of the $CP$ asymmetry drops to less
than $1\%$ while still leaving open the possibility of higher values
(of order $5\%$) in the low branching ratio region.  This implies that
in the presence of a cancellation mechanism to satisfy EDM constraints
large asymmetries are still possible in this decay. In this regard,
there was recently some controversy on the possible size of this
asymmetry with EDM constraints \cite{aoki-cho,goto}.  In fact both
works assumed that $\phi_A$ was basically unconstrained by EDM
experiments. In this conditions Ref.~\cite{aoki-cho} found asymmetries
very similar to our result. However, it was then pointed out
\cite{goto} that RGE effects tend to reduce the phase of $A_t$ at the
electroweak scale and the asymmetry was again reduced below $2\%$. As
we have shown here, having a non--vanishing $\phi_\mu$ through a
cancellation mechanism can, in some cases, enhance the asymmetry
around $5\%$ in the low branching ratio region.

Before concluding this subsection, we would like to comment on the
issue regarding the sign of $C_7^{eff} (m_b)$. The $b\rightarrow s
\gamma$ rate constrains the absolute value of this WC
and it does not give any information on its sign which, on the other
hand, has a strong impact on inclusive and exclusive $b\rightarrow s
\ell^+ \ell^-$ transitions.  Here we only want to discuss the possibility
to achieve the sign flip in the class of models we consider and we
refer the reader to Refs.~\cite{cho,handoko,ignazio} for a discussion
of positive $C_7$ phenomenology.  Our result is that no points with
$\hbox{Re} (C_7^{eff}) >0$ survive after imposing the EDM's
constraints on the phases\footnote{If the EDM constraints are not
imposed, it is possible to obtain points with $\hbox{Re} (C_7^{eff})
>0$ but always with a large $\hbox{Im} (C_7^{eff})$.}.  This
conclusion is partially due to the presence of a correlation between
the charged Higgs, light stop and light chargino masses.  In fact, in
order to get a large positive chargino contribution it is required to
have $\mu>0$, large $\tan \beta$, chargino and stop masses as light as
possible.  This, in turn, implies a relatively light charged Higgs so
that its contribution, which is always negative, tends to balance the
chargino contribution preventing the sign flip.  We must say that such
a conclusion could be modified if the GUT scale conditions we impose
are significantly relaxed.

\subsection{Unitarity triangle}

$CP$ violation in the SM is completely encoded in the CKM mixing
matrix.  Thanks to unitarity of this matrix, the existence of $CP$
violation in the SM is equivalent to the presence of a non--trivial
unitarity triangle. Therefore, the measure of the unitarity triangle
is a direct test of the CKM origin and possible new sources of $CP$
violation. The best triangle for this purpose is the triangle produced
by the product of the first and third columns of the CKM mixing
matrix. In this triangle all three sides are of third order in the
Wolfenstein parameter $\lambda$ and normalizing the three sides with
respect to the base $K_{cd} K_{cb}^*$ we obtain,
\begin{equation}
\label{utnorm}
\Frac{K_{ud} K_{ub}^*}{K_{cd} K_{cb}^*} + 1 +\Frac{ K_{td} K_{tb}^*}
{K_{cd} K_{cb}^*} = 0,
\end{equation}
In fact, the shape and size of this triangle is overconstrained by
many different $CP$--violating and $CP$--conserving experimental
observables.  In this regard, some observables, being tree--level
contributions in the SM, are basically unaffected by new physics
contributions, as for instance the values of $|K_{cb}|$ and $|K_{ub}|$
which basically determine the size of one of the sides of the
triangle.

In first place, from semileptonic decays of the $B$ meson we have a
direct measurement of $|K_{ub}/K_{cb}|=0.093 \pm 0.018$
\cite{PDG}. This implies,
\begin{equation}
\label{rad}
\left|\Frac{K_{ud} K_{ub}^*}{K_{cd} K_{cb}^*}\right| = \frac{1}{\lambda}
\left|\Frac{K_{ub}}{K_{cb}}\right| = 0.42 \pm 0.08
\end{equation}
In the $(\rho,\eta)$ plane this constraint is represented as a circle centered
in $(\rho,\eta) = (0,0)$ with a radius given in Eq~(\ref{rad}).

All other observables used to constrain this triangle are already
present at 1--loop in the SM and hence can be affected by the
inclusion of additional contributions from SUSY. For instance, the
third side, determined by $|K_{td}|$ is measured only indirectly
through $B^0_d$--$\bar{B}^0_d$ or $B^0_s$--$\bar{B}^0_s$ mixing which
in principle, can receive sizeable contributions from SUSY loops,
modifying the SM determination of this side. The main constraint on
the unitarity triangle is provided by the observation of indirect $CP$
violation in the neutral $K$ system, namely $\varepsilon_K$. This
measurement implies, in the SM, that the unitarity triangle does not
collapse to a line and there is an observable phase in the CKM
matrix. In the presence of SUSY the existence of a non--zero
$\varepsilon_K$ does not necessarily require the presence of a phase
in the CKM matrix, i.e. a non-trivial unitarity triangle
\cite{gabbiani,KvsB,piai}.
Still, as shown in \cite{flavor}, in a flavor blind MSSM,
$\varepsilon_K$ is always proportional to the phase in the CKM matrix
and hence a non--trivial triangle is also required. However, the shape of
this triangle is always modified by the new SUSY contributions, and a new
fit of this triangle is required \cite{ali-london,buras,branco}.
All these measurements allow, with the present experimental data, a good
determination of the unitarity triangle within a well defined model.
Nevertheless, the recent arrival of new data from the $B$--factories provide
independent information on this triangle. In particular, the $B^0$ $CP$
asymmetries measure directly the internal angles in this triangle.

Hence, to perform a complete fit of the unitarity triangle in a general
flavor blind MSSM, we must include new contributions from sfermion loops
and charged Higgs.
Both $B^0_d$--$\bar{B}^0_d$, $B^0_s$--$\bar{B}^0_s$  mixings and
$\varepsilon_K$ are fully described by the $\Delta F=2$ effective Hamiltonian,
${\cal{H}}_{eff}^{\Delta F=2}$. In this framework, the only four quark
operators are, neglecting light quark masses,
\begin{eqnarray}
\label{DF=2}
&{\cal{H}}_{eff}^{\Delta F=2}=-\Frac{G_{F}^{2} m_{W}^{2}}{(2 \pi)^{2}}
(K_{td}^{*} K_{tq})^{2}\ \left( \cal{C}_{1}(Q)\  \bar{d}^{\alpha}_{L}\gamma^{Q}
q^{\alpha}_{L}\cdot \bar{d}^{\beta}_{L}\gamma_{Q}q^{\beta}_{L}
\ +\  \cal{C}_{2}(Q)\  \bar{d}^{\alpha}_{L}q^{\alpha}_{R}\cdot \bar{d}^{\beta}_{L}
q^{\beta}_{R}\right. \nonumber \\
&\left. +\ \cal{C}_3(Q)\ \bar{d}^{\alpha}_{L}q^{\beta}_{R}\cdot
\bar{d}^{\beta}_{L}q^{\alpha}_{R}\right)
\end{eqnarray}
with $q=s, b$ for the $K$ and $B$--systems respectively and $\alpha,
\beta$ are color indices. The value of the WCs at $m_W$ is,
\begin{eqnarray}
\label{wilson1}
{\cal{C}_{1}}(m_W)&=&{\cal{C}_{1}^{W}}(m_W)+{\cal{C}_{1}^{H}}(m_W)+
{\cal{C}_{1}^{\chi}}(m_W) \\
\cal{C}_1^W (m_W) &=& \sum_{\alpha \gamma} \Frac{K_{\alpha d}^{*}
K_{\alpha q}K_{\gamma d}^{*} K_{\gamma q}}{(K_{td}^{*} K_{tq})^2}\
G (x_{\alpha}, x_{\gamma}) \nonumber \\
\cal{C}_1^{H^+} (m_W) &=& \sum_{\alpha \gamma} \Frac{K_{\alpha d}^{*}
K_{\alpha q}K_{\gamma d}^{*} K_{\gamma q}}{(K_{td}^{*} K_{tq})^2}
\left[\Frac{1}{4\ \tan^4 \beta}\ x_\alpha x_\gamma  Y_1(x_{H}, x_{H},
x_\alpha, x_\gamma)+\right.  \nonumber \\
&&\left. \Frac{1}{2\ \tan^2 \beta}\ x_\alpha x_\gamma  Y_1(1, x_{H},
x_\alpha, x_\gamma) - \Frac{2}{\tan^2 \beta}\sqrt{x_\alpha x_\gamma}\
Y_2(1, x_{H}, x_\alpha, x_\gamma)\right] \nonumber\\
\cal{C}_1^\chi (m_W) &=& \sum_{i,j=1}^{2} \sum_{k, l=1}^{6}
\sum_{\alpha \gamma \alpha^\prime \gamma^\prime} \Frac{K_{\alpha^\prime d}^{*}
K_{\alpha q}K_{\gamma^\prime d}^{*} K_{\gamma q}}{(K_{td}^{*} K_{tq})^2} 
G^{(\alpha,k)i} G^{(\alpha^\prime,k)j*} G^{(\gamma^\prime,l)i*}
G^{(\gamma,l)j}\  Y_1(z_{k}, z_{l}, s_i, s_j)\nonumber
\end{eqnarray}
\begin{eqnarray}
\cal{C}_2^{H^+} (m_W) &=& \sum_{\alpha \gamma} \Frac{K_{\alpha d}^{*}
K_{\alpha q}K_{\gamma d}^{*} K_{\gamma q}}{(K_{td}^{*} K_{tq})^2}\ \Frac{m_q^2}
{m_W^2}\sqrt{x_\alpha x_\gamma}\
Y_2(x_H, x_{H}, x_\alpha, x_\gamma)] \\
\cal{C}_3^\chi (m_W) &= &\sum_{i,j=1}^{2} \sum_{k, l=1}^{6} \sum_{\alpha \gamma
\alpha^\prime \gamma^\prime}
\Frac{K_{\alpha^\prime d}^{*} K_{\alpha q}K_{\gamma^\prime d}^{*}
K_{\gamma q}}{(K_{td}^{*} K_{tq})^2}\  \Frac{m_q^2}{2 m_W^2 \cos^2 \beta} \times
\nonumber \\
&&H^{(\alpha,k)i} G^{(\alpha^\prime,k)j*} G^{(\gamma^\prime,l)i*}
H^{(\gamma,l)j} Y_2(z_k, z_l, s_i, s_j)
\label{wilson3}
\end{eqnarray}
where all the WCs are evaluated
at $m_W$, and we have, $x_\alpha= m_\alpha^2/m_W^2$,
$z_k = m_{\tilde{u}_k}^2/m_W^2$ and $s_i =m_{\tilde{\chi}_i}^2/m_W^2$.
The explicit expressions for the loop functions can be found in the Appendix.

We must remember that in any flavor blind scenario a LR transition
must always go through a Yukawa coupling and given that the
right--handed mixing can always be rotated away, these LR transition
are always associated with the Yukawa coupling of the right handed
fermion\footnote{Notice that this is not true anymore if a new
right--handed coupling is present, as it is the case in general
non--universal MSSM model \cite{EDMfree,piai,tatsuo,newflavor}}.
Hence, the $\cal{C}_2$ and $\cal{C}_3$ WC are suppressed by
$m_q^2/m_W^2$ and $m_q^2/(m_W^2 \cos^2 \beta)$ respectively.  Then, it
is easy to understand that the main four--fermion operator in our
model, as well as in the SM, will always be the first operator in
Eq.~\ref{DF=2}, $Q_{1}$, that involves only left--handed quarks. In
fact, the remaining operators can be neglected in $K$--$\bar{K}$
mixing due to the smallness of $m_s$.  In the $B$ system with large
$\tan \beta$ these operators are not suppressed in principle, however,
in this case, the $b\to s \gamma$ branching ratio strongly constrains
these contributions and as a result in $B$--$\bar{B}$ mixing they can
also be neglected.  Moreover, in the limit of vanishing
intergenerational mixing in the sfermion mass matrices the $\cal{C}_1$
WC is real in very good approximation
\cite{fully,CPcons}.
Hence, using Eqs~(\ref{wilson1}--\ref{wilson3}) we can calculate
$\varepsilon_K$, $B_d$--$\bar{B}_d$ and $B_s$--$\bar{B}_s$ mixing as,
\begin{eqnarray}
\label{epsK}
\varepsilon_K\ &=&\ \Frac{G_F^2 m_W^2}{12 \pi^2 \sqrt{2}\ \Delta
M_K}\ f_K^2\  M_K\ B^K_1\ \mbox{Im} \left\{(K_{cd}^* K_{cs})^2\ \eta_{c}
\ G_1(x_c,x_c) + \right. \nonumber \\
&&\left. (K_{cd}^* K_{cs} K_{td}^* K_{ts})\ \eta_{tc}\ G_1(x_t,x_c) +
(K_{td}^* K_{ts})^2\ \eta_{t} \ \left( G_1(x_t,x_t) + \cal{C}^\chi_1 +\cal{C}_1^{H^+}
\right) \right\}
\end{eqnarray}
where we use $B^K_1=0.94 \pm 0.15$ and $\eta_t=0.574$, $\eta_c=1.38$,
$\eta_{tc}=0.47$\cite{ali-london}.
\begin{eqnarray}
\label{DMbd}
\Delta M_{B_d}\ &=&\ \Frac{G_F^2 m_W^2}{6 \pi^2}\ f_{B_d}^2\ B_{B_d}\ M_{B_d}\
\ \eta_1\ (K_{td} K_{tb})^2\ \cal{C}^{(B_d)}_1 \\
\label{DMbs}
\Frac{\Delta M_{B_s}}{\Delta M_{B_d}} &=&\ \xi^2 \ \Frac{M_{B_s}}{M_{B_d}}\ 
\Frac{(K_{ts})^2\ \cal{C}^{(B_s)}_1}{(K_{td})^2\ \cal{C}^{(B_d)}_1}
\end{eqnarray}
with $\sqrt{f_{B_d}^2 B_{B_d}} = (230 \pm 40)\ \mbox{MeV}$, $\eta_1= 0.55$, and
$\xi =\sqrt{f_{B_s}^2 B_{B_s}}/\sqrt{f_{B_d}^2 B_{B_d}} = 1.16 \pm 0.05$.
The experimental values for these observables are,
$\varepsilon_K= (2.28 \pm 0.05) \times 10^{-3}$,
$\Delta M_{B_d}= 0.487 \pm 0.014\ ps^{-1}$ and 
$\Delta M_{B_s}\geq 15.0\ ps^{-1}$ at $95 \%$ C.L..
Fixing all the SUSY 
and hadronic
parameters and expressing these constraints as functions of the CKM
parameters $(\rho,\eta)$, $\Delta M_{B_d}$ gives rise 
to a circle in the $\rho$--$\eta$ plane centered in $(1,0)$ and 
similarly, $\varepsilon_K$ specifies an hyperbola. 
The $\Delta M_{B_s}$ constraint is approximately a circle centered
as well in $(1,0)$.

\begin{figure}[H]
\begin{center}
\epsfig{file=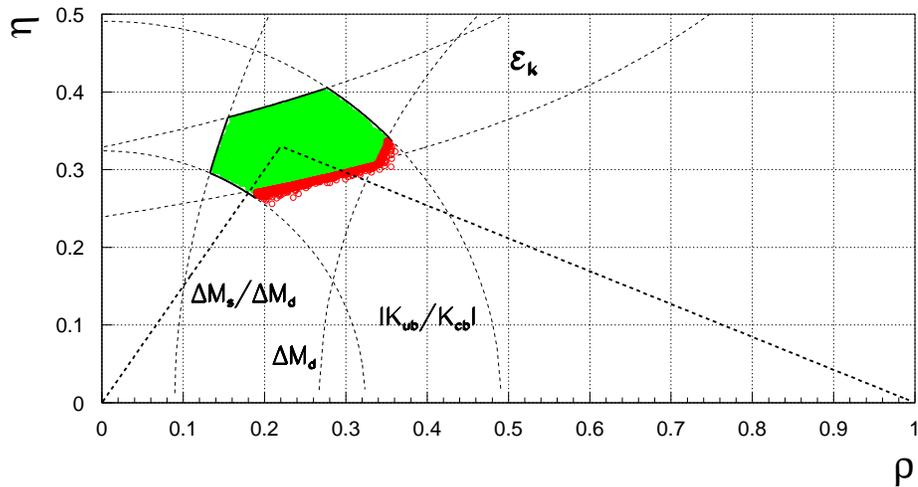,width=0.80 \linewidth}
\caption{Fit of the unitary triangle in the flavor blind MSSM framework.
The filled region corresponds to the SM fit and the open circles
show the possible deviations that can occurr in these models.}
\label{fig:unitarity}
\end{center}
\end{figure}

In Figure~\ref{fig:unitarity} we present the allowed $(\rho,\eta)$ range
in the flavor blind MSSM as described in the previous sections.
Here, the grey area corresponds to the region already allowed in the SM
while the open circles present the deviation induced by the SUSY
contributions. Clearly, under these conditions, no large deviations from the
SM predictions can be expected. In fact, the relative heaviness of the
SUSY spectrum and smallness of mixing angles restricts the effects to a small
region below the SM area. However, it is interesting to notice that,
due to the fact that the SUSY contributions are always proportional
to a CKM element and interfere constructively with the SM, the value of
$\beta$ tends to be reduced, in the direction of the recent experimental
measurements at B factories.

This result differs from similar analyses already present in the literature
under the name of Minimal Flavor Violation MSSM \cite{ali-london,buras,buras2}.
In these works they assume that the only flavor structure in the model is the
CKM mixing matrix and they consider only chargino--stop, charged Higgs and W
contributions. With these conditions, they are able to find very large 
deviations
in $\varepsilon_K$ and $\Delta M_{B_d}$. The main difference with these works
is that they consider SUSY masses and mixings as independent variables
constrained by low energy experiments. Our framework is much more restrictive
and, as shown in section~\ref{section:spectrum}, the RGE
evolution implies that the lightest stop is $250\ \mbox{GeV}$ for a chargino 
of $100\ \mbox{GeV}$ and the lightest charged Higgs is $300\ \mbox{GeV}$. 
This can be compared
with $m_{\chi^\pm}=m_{\tilde{t}_1}=m_{H^\pm}= 100\ \mbox{GeV}$ used in 
\cite{ali-london} or
$m_{\chi^\pm}>90\ \mbox{GeV}$, $m_{\tilde{t}_1}> 90\ \mbox{GeV}$ and $m_{H^\pm}> 100\ \mbox{GeV}$
in \cite{buras}. Similarly, these papers take the mixing angles as completely
free while in our flavor blind scenario, as we can see in Eq.~(\ref{stopmix}),
the value of the stop and chargino mixing angles are determined in terms of 
the same $M_{GUT}$ inputs.  These facts
forbid supersymmetric contributions to compete with SM loops and consequently
only small deviations from the SM range are allowed.

\subsection{Muon anomalous magnetic moment}

In this subsection, we analyze the impact of the recent Brookhaven E821
measurement of the positive muon anomalous magnetic moment on our
analysis. The actual world average for this quantity~\cite{brook} and
the corresponding SM prediction~\cite{marciano} are
\begin{eqnarray}
a_{\mu^+}(\hbox{exp}) &=& 11659203(15) \times 10^{-10}, \nonumber\\
a_{\mu^+}(\hbox{SM}) &=& 11659160(7) \times 10^{-10}, \nonumber
\end{eqnarray}
so that the difference,
\begin{equation}
\delta a_{\mu^+} = +43(16) \times 10^{-10},
\label{deltaamu}
\end{equation}
gives a $2.6 \; \sigma$ deviation from the SM.

The SM estimate given above is based on a recent computation by Davier
and H\"{o}cker~\cite{davier} which is considered to be the most
precise published analysis to date~\cite{marciano1}.  The bulk of the
theoretical error is due to the hadronic contribution which is
obtained from $\sigma (e^+ e^- \rightarrow \mbox{hadrons})$ via a
dispersion relation.  In order to minimize the errors, Davier and
H\"{o}cker supplemented the $e^+ e^- \rightarrow \pi^+ \pi^-$ cross
section using data from tau decays.  The issue whether or not to use
these decays was extensively debated in the literature. On this basis,
the author of Ref.~\cite{yndurain} questions the superiority of the Davier
and H\"{o}cker
analysis and concludes, after a survey of all the SM theoretical
predictions for $a_{\mu^+}$, that it is definitely too early to
advertise any deviation from the SM.  In the next few years, the
situation will become clearer because the experimental uncertainty on
$a_{\mu^+}$ will be reduced and new data on the $e^+ e^- \rightarrow
\pi^+ \pi^-$ cross section will be taken.

In the following analysis, we adopt the estimate (\ref{deltaamu})
in order to understand its phenomenological impact on our flavor blind
MSSM in case this mismatch is confirmed in the future.

In SUSY theories, $a_{\mu^+}$ receives contributions via vertex diagrams with
$\chi^0$--$\tilde \mu$ and $\chi^\pm$--$\tilde \nu$ 
loops~\cite{moroi,nath,everett,gm2pap,gondolo,nath2,moroi2,ellisgm2,martin,ellisgm21,baer}.
The chargino diagram strongly dominates in almost all the parameter
space. For simplicity, we will present here only the dominant part of
chargino contribution (the complete expressions that we use in the
numerical simulation can be found in Ref.~\cite{moroi}, see also
Ref.~\cite{nath} for a discussion on $CP$ violating phases):
\begin{equation}
\delta a_{\mu^+}^{\chi \tilde\nu} \simeq -{g_2^2\over 8 \pi^2}
{m_\mu^2\over m_{\tilde\nu}^2}
 \sum_{i=1}^2
 \frac{m_{\chi_i} \hbox{Re} (U_{i2} V_{i1})}
      {\sqrt{2} M_W \cos \beta}
 F_3 \left({m_{\chi_i}^2\over m_{\tilde\nu}^2}\right)
\label{gm2char}
\end{equation}
where the loop function $F_3$ is given in the appendices.
The most relevant feature of Eq.~(\ref{gm2char}) is that the sign of
$\delta a_{\mu^+}^{\chi \tilde\nu}$ is fixed by
$\hbox{sign}[\hbox{Re}(U_{12} V_{11}) ] = - \hbox{sign}[\hbox{Re}(\mu)]$.
Comparison with Eq.~(\ref{deltaamu}) implies that the new Brookhaven result
strongly favors the $\mu>0$ region in a MSSM scenario. Indeed, this
has very important consequences in the observables analyzed in the
previous sections.

\begin{figure}[H]
\begin{center}
\epsfig{file=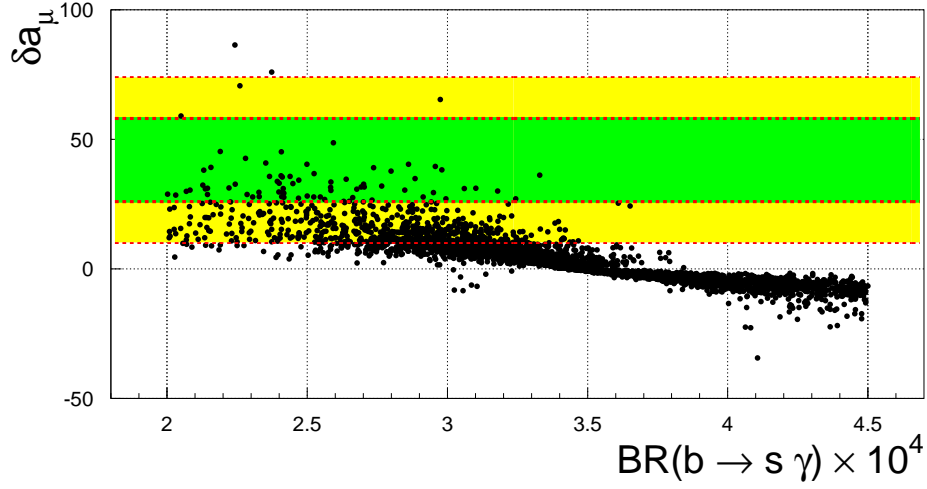,width=0.8 \linewidth}
\caption{Plot of the SUSY contribution to $a_{\mu^+}$ versus the branching
ratio of $b\rightarrow s \gamma$. The bands represent the $1\sigma$
and $2\sigma$ allowed regions according to Eq.~(\ref{deltaamu}).}
\label{fig:amu}
\end{center}
\end{figure}

\begin{figure}[H]
\begin{center}
\epsfig{file=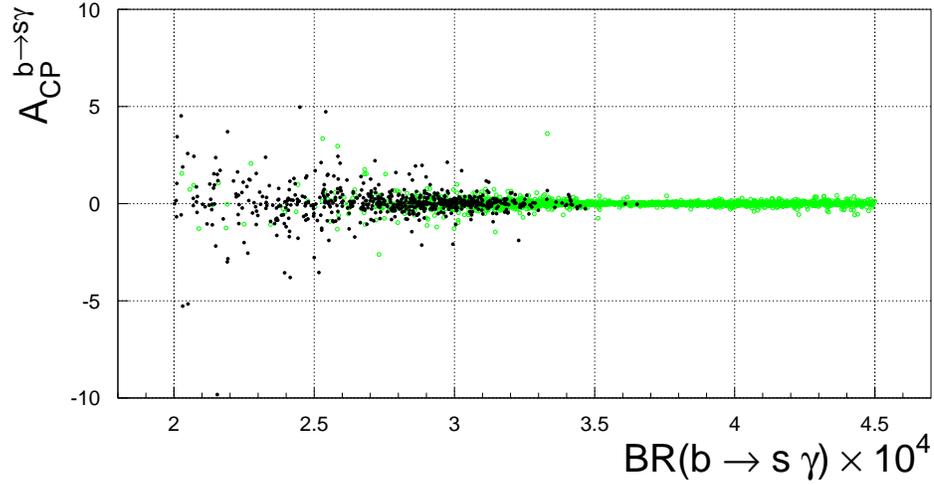,width=0.8 \linewidth}
\caption{Plot of the $CP$ asymmetry versus the branching ratio of
$b\rightarrow s \gamma$. We allow only for points whose phases satisfy
the EDM's constraints.  The black dots satisfy the 2
sigma bound implied by Eq.~(\ref{deltaamu}).}
\label{fig:acpamu}
\end{center}
\end{figure}

In first place, recalling our previous discussion of the $b\rightarrow
s \gamma$ decay, this implies that the chargino contribution to the
$C_7$ WC is preferably positive. Therefore, it interferes
destructively with the SM term and then the BR is generally smaller
than the SM value. In this situation, we expect higher values of the
$CP$ asymmetry to be slightly favored too.  These qualitative arguments
are confirmed by the numerical analysis that we summarize in
Fig.~\ref{fig:amu} and \ref{fig:acpamu}.  In Fig.~\ref{fig:amu}, we
present the correlation among $\delta a_\mu$ and $\mbox{BR}( b \to s
\gamma)$, together with the $1 \sigma$ and $2\sigma$ preferred ranges
for $\delta a_\mu$. In this plot, we can see that the required
contribution in $\delta a_\mu$ implies a low branching ratio in the $b
\to s \gamma$ decay. Similarly, in Fig.~\ref{fig:acpamu}, we show as
black dots the points of the parameter space that reproduce the
measured anomalous magnetic moment and as open circles all other
points. Then we have that, in the presence of sizeable SUSY phases,
large values of the $CP$ asymmetry can also be expected.

\begin{figure}[H]
\begin{center}
\epsfig{file=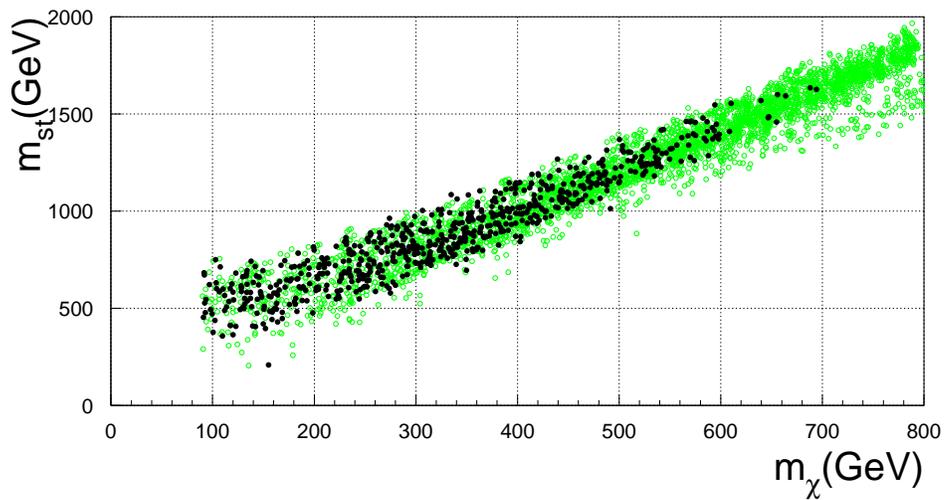,width=0.8 \linewidth}
\caption{Impact of the $a_{\mu^+}$ constraint on the lightest chargino
and stop masses. The black dots satisfy the 2
sigma bound implied by Eq.~(\ref{deltaamu}).}
\label{fig:mchmstgm2}
\end{center}
\end{figure}

Finally, in Fig.~\ref{fig:mchmstgm2}, we plot once more the
correlation between the lightest chargino and stop masses to see the
impact of the $a_{\mu^+}$ constraint on them. Here, black dots are the
points of the parameter space that reproduce the required value of
$\delta a_\mu$.  We confirm the presence of an upper bound on the
chargino mass of about $700\ \mbox{GeV}$ for very large $\tan \beta$ (of order 50)
\cite{gm2pap}, and lower for smaller values of
$\tan \beta$.  This bound is essentially due to our assumption of
gaugino mass unification. In fact, as was recently pointed out in
Ref.~\cite{martin}, if this hypothesis is relaxed, the chargino and
neutralino masses are uncorrelated and in the large chargino mass
region $a_{\mu^+}^{\chi^0 \tilde\mu}$ can compensate (for a
sufficiently light smuon) the exceedingly small $a_{\mu^+}^{\chi^\pm
\tilde \nu}$. In this way, we obtain a big enough contribution to the
anomalous magnetic moment of the muon even in the limit of decoupling
chargino sector. Still, we would like to stress here that in any RGE
evolved MSSM from a GUT scale with gaugino mass unification this
measurement has important consequences on the complete MSSM
spectrum. In particular, as shown here, the chargino--stop mass
correlation implies, with no additional restriction on the SUSY
parameter space, the presence of an upper bound on the light stop mass
of $m_{\tilde{t}} \leq 1500\ \mbox{GeV}$.

\section{Conclusions and general results}


In this paper, we analyze the low energy phenomenology of {\it flavor
blind MSSMs}, and in particular we focus on $CP$ violating
observables.  We calculate the MSSM spectrum at the electroweak scale
using two--loop RGEs in terms of the initial conditions at the GUT
scale.  We apply the constraints from direct searches, the
$\rho$--parameter, the absence of charge and color breaking minima and
the requirement of the lightest SUSY particle (LSP) to be neutral.
Using the points that survive these bounds, we study the further
restrictions on the SUSY parameters, especially the complex phases,
derived from the electron EDM and the $b \to s \gamma$ decay.

In the resulting allowed regions of the parameter space, we analyze
the predictions of these models on $\epsilon_K$, $\Delta M_{B_d}$ and
$\Delta M_{B_s}$ as well as on the $b \rightarrow s \gamma$ $CP$
asymmetry and on the muon anomalous magnetic moment, whose relevance
was strengthened by some recent experimental results.

The well known gluino dominance on the RG evolution gives rise to
strong correlations between gaugino masses, squark masses and mixing
angles at the $m_W$ scale.  In particular, only a narrow band in the
stop--chargino mass plane is allowed: the current lower bounds on the
chargino mass imply that the lightest stop must be heavier than about
250 GeV.  The charged Higgs boson mass is generally above 400 GeV
although it is possible to find lighter masses at moderate values of
$\tan \beta$.  The impact of the electron EDM results mainly in a
strong constraint on $\phi_\mu$, while $\phi_A$ can reach much higher
values.  Taking into account the possibility of cancellations between
different SUSY contributions, we find values of $\phi_\mu$ up to 0.4
and $\phi_A$ is essentially unconstrained.  The $b\rightarrow s
\gamma$ constraint cuts a sizable region with large $\tan \beta$ and
light scalar and gaugino masses and hence plays an important role in
the determination of the finally allowed parameter space.

Taking into account all these results we calculate the $b\rightarrow s
\gamma$ $CP$ asymmetry which turns out to reach values up to $5 \; \%$
for relatively small values of $BR(b\rightarrow s \gamma)$.  Such
asymmetries are within the reach of the current B--factory experiments
and therefore they are a very useful tool to check the EDM
cancellation mechanism. In fact, if large phases survive the EDM
constraints via this mechanism, the $CP$ asymmetry can reach the above
upper limit.  Concerning the impact of flavor blind SUSY on the
unitarity triangle fit, we do not find any sizeable deviation from the
SM allowed region: this is due to the relative heaviness of the SUSY
spectrum, especially of the top squark, lightest chargino and charged
Higgs boson.  Finally we focus on possible large SUSY effects on the
muon anomalous magnetic moment. It is possible to match the recent
experimental determination for positive $\mbox{Re} (\mu)$ values which
are also favored by the branching ratio of $b\rightarrow s
\gamma$. In fact, the points that reproduce the experimental value of
$a_{\mu^+}$, can have, at the same time, a large $CP$ asymmetry in the
$b\rightarrow s \gamma$ decay. This measurement also implies an upper
bound on the chargino and stop masses respectively of 700 GeV and 1500
GeV.

In summary, these flavor blind MSSMs have a small impact on $\Delta F
= 2$ observables and hence do not modify sizeably the SM fits of the
unitarity triangle. On the other hand, the EDMs and the $CP$ asymmetry
in $b\rightarrow s \gamma$ are closely correlated. If a
$\cal{A}^{CP}_{b\to s \gamma}$ is observed, the only possibility to
account for it in a flavor blind SUSY context is that large
cancellations among SUSY contributions in the EDMs occur.  In
conclusion, the EDM, the $CP$ asymmetry in $b\to s \gamma$ and the
anomalous magnetic moment of the muon constitute possible candidates
for significant deviations from the SM expectations even if the
breaking of supersymmetry has nothing to do with the origin of
the flavor in the theory.

\section*{Acknowledgements}
Two of us (E.L. and A.M.) thank A. Ali and S. Bertolini for
interesting discussions. We acknowledge N. Fornengo and P. Ullio for
useful comments on dark matter constraints.  This work was supported
by the `Fonds zur F\"orderung der wissenschaftlichen Forschung' of
Austria, FWF Project No.  P13139-PHY, by the EU TMR Network Contracts
No. HPRN-CT-2000-00148, HPRN-CT-2000-00149 and HPRN-CT-2000-00152 and
by Cooperazione scientifica e tecnologica Italia-Austria 1999-2000,
Project No. 2; W.P. was supported by the Spanish Ministry of Education
and Culture under the contract SB97-BU0475382 and by DGICYT grant
PB98-0693; T.G. acknowledges financial support from the DOE grant
DE-FG02-96ER40967; E.L. and O.V. thank SISSA for support during the
first stages of this work. E.L. was supported by the Alexander Von
Humboldt Foundation; O.V. acknowledges financial support from a Marie
Curie E.C. grant (HPMF-CT-2000-00457) and partial support from Spanish
CICYT AEN-99/0692.

\appendix

%
\section{Sfermion Mass Matrix}

The $6\times 6$ sfermion mass matrices are given by
\begin{equation}
M^{2}_{\tilde{f}}
=
\left(\begin{array}{cc}
  M^{2}_{\tilde{f}_L}
+ ( T^{3}_{I} - Q_{f} \sin^{2}\theta_{W} ) \cos 2\beta \, m_{Z}^{2} 
+ M_{f}^{2}
 & Y_{A_f}^*  \displaystyle {v \over \sqrt{2}}\ \Omega (\beta) - 
M_{f}\mu \Theta(\beta) 
\\
Y_{A_f} \displaystyle {v \over \sqrt{2}}\ \Omega (\beta) - 
M_{f}\mu^{*} \Theta(\beta) 
 & M^{2}_{\tilde{f}_R}
+ Q_{f} \sin^{2}\theta_{W} \cos 2\beta \, m_{Z}^{2}
+ M_{f}^{2}
\end{array}\right)
\enspace ,
\end{equation}
where 
\begin{eqnarray}
\cases{\Theta(\beta)= \cot\beta,\ \Omega(\beta)=\sin\beta &
 for $T^{3}_{I} =  \frac{1}{2}$ \cr
                    \Theta(\beta)= \tan\beta,\ \Omega(\beta)=\cos\beta & 
for $T^{3}_{I} = -\frac{1}{2}$ \cr}
\enspace ,
\end{eqnarray}
and $Y_{A_f}$ are the trilinear matrices equal at $M_{GUT}$ to 
$Y_{A_f}= Y_f A_f$.
These matrices are diagonalized by the $6\times 6$ unitary matrices 
$\Gamma_{f}$:
\begin{equation}
\mbox{diag} (M_{\tilde f_1},\dots,M_{\tilde f_6}) =
\Gamma_{\tilde f}^{} \cdot M_{\tilde f}^2 \cdot \Gamma_{\tilde f}^\dagger \enspace . 
\end{equation}
The $6\times 3$ left and right block components of the mixing matrices are
defined as:
\begin{equation}
\Gamma_{\tilde f}^{6\times 6} = \left( \Gamma^{6\times 3}_{\tilde f L}  \ \ 
      \Gamma^{6\times 3}_{\tilde f R} \right) \enspace .
\end{equation}

In the flavor blind scenario, the most important off--diagonal entry
in the above squared mass matrices is the third generation LR   
mixing. Below we present the analytic expressions of the $2\times 2$
stop system: 
\begin{equation}
M^{2}_{\tilde{t}}
=
\left(\begin{array}{cc}
  M^{2}_{\tilde{t}_{LL}}
 & e^{-i \varphi_{\tilde{t}}} M^{2}_{\tilde{t}_{LR}}
\\
  e^{i \varphi_{\tilde{t}}} M^{2}_{\tilde{t}_{LR}}
 & M^{2}_{\tilde{t}_{RR}}
\end{array}\right)
\enspace ,
\label{s1}
\end{equation}
where
\begin{eqnarray}
  M^{2}_{\tilde{t}_{LL}}
&=&
  m^{2}_{Q_3}
+ ( {1\over 2} - {2\over 3} \sin^{2}\theta_{W} ) \cos 2\beta \, m_{Z}^{2}
+ m_{t}^{2}
\enspace ,
\\
  M^{2}_{\tilde{t}_{RR}}
&=&
  m^{2}_{U_3}
+ {2\over 3} \sin^{2}\theta_{W} \cos 2\beta \, m_{Z}^{2}
+ m_{t}^{2}
\enspace ,
\\
  M^{2}_{\tilde{t}_{LR}}
&=&
  m_{t} | A_{t} - \mu^{*} \Theta(\beta) |
\enspace ,
\\
\label{sfphase}
  \varphi_{\tilde{t}}
&=&
  \arg [ A_{t} - \mu^{*} \Theta(\beta) ]
\enspace ,
\end{eqnarray}
The eigenvalues are given by
\begin{equation}
2 m^{2}_{\tilde t_1, \tilde t_2}
= ( M^{2}_{\tilde{t}_{LL}} + M^{2}_{\tilde{t}_{RR}} )
\mp \sqrt{ ( M^{2}_{\tilde{t}_{LL}} - M^{2}_{\tilde{t}_{RR}} )^{2}
         + 4 ( M^{2}_{\tilde{t}_{LR}} )^{2}}
\enspace ,
\end{equation}
with $m^2_{\tilde t_1} \le m^{2}_{\tilde t_2}$.
We parametrize the mixing matrix ${\mathcal R}^{\tilde{t}}$ so that
\begin{equation}
\label{s2}
\left(\begin{array}{c}
  \tilde{t}_{1} \\ \tilde{t}_{2}
\end{array}\right)
=
{\mathcal R}^{\tilde{t}}
\left(\begin{array}{c}
  \tilde{t}_{L} \\ \tilde{t}_{R}
\end{array}\right)
=
\left(\begin{array}{cc}
  e^{\frac{i}{2} \varphi_{\tilde{t}}} \cos \theta_{\tilde{t}}
 & e^{-\frac{i}{2} \varphi_{\tilde{t}}} \sin \theta_{\tilde{t}}
\\
  - e^{\frac{i}{2} \varphi_{\tilde{t}}} \sin \theta_{\tilde{t}}
 & e^{-\frac{i}{2} \varphi_{\tilde{t}}} \cos \theta_{\tilde{t}}
\end{array}\right)
\left(\begin{array}{c}
  \tilde{t}_{L} \\ \tilde{t}_{R}
\end{array}\right)
\enspace ,
\end{equation}
where $\varphi_{\tilde{t}}$ is given in Eq.~(\ref{sfphase}) and
\begin{eqnarray} &&
\cos\theta_{\tilde{t}}
=
\frac{-M^{2}_{\tilde{t}_{LR}}}{\Delta}
\leq 0
\enspace , \quad
\sin\theta_{\tilde{t}}
=
\frac{M^{2}_{\tilde{t}_{LL}} - m^2_{\tilde t_1}}{\Delta}
\geq 0
\enspace ,
\nonumber \\ &&
\Delta^{2}
=
  ( M^{2}_{\tilde{t}_{LR}} )^{2}
+ ( m^2_{\tilde t_1} - M^{2}_{\tilde{t}_{LL}} )^{2}
\enspace .
\end{eqnarray}


\section{Chargino Mass Matrix}

The chargino mass matrix
\begin{equation}\label{charmass}
M^{\tilde{\chi}^{+}}_{\alpha\beta} =
\left(
\begin{array}{cc}
  M_2                        & m_{W} \sqrt{2} \sin\beta  \\
  m_{W} \sqrt{2} \cos\beta & \mu
\end{array}
\right)
\end{equation}
can be diagonalized by the biunitary transformation
\begin{equation}
U^{*}_{j\alpha} M^{\tilde{\chi}^{+}}_{\alpha\beta} V^{*}_{k\beta}
= m_{\tilde{\chi}_{j}^{+}} \delta_{jk}
\enspace ,
\end{equation}
where $U$ and $V$ are unitary matrices such that
$m_{\tilde{\chi}_{j}^{+}}$ are positive and
$m_{\tilde{\chi}_{1}^{+}} < m_{\tilde{\chi}_{2}^{+}}$.


\section{Neutralino Mass Matrix}

We define $N_{\alpha j}$ as the unitary matrix which makes the complex
symmetric neutralino mass matrix diagonal with positive diagonal
elements:
\begin{equation}
N_{\alpha j} M^{\tilde{\chi}^{0}}_{\alpha\beta} N_{\beta k}
= m_{\tilde{\chi}^{0}_{j}}\delta_{jk}
\enspace ,
\end{equation}
where $m_{\tilde{\chi}^{0}_{j}} < m_{\tilde{\chi}^{0}_{k}}$ for $j<k$.
In the basis \cite{Bartl:1989ms}:
\begin{equation}
\psi_{\alpha} =
\{ - i\tilde{\gamma},-i\tilde{Z},\tilde{H}^{a},\tilde{H}^{b} \}
\enspace ,
\end{equation}
the complex symmetric neutralino mass matrix has the form
\begin{equation}\label{neutmass}
M^{\tilde{\chi}^{0}}_{\alpha\beta} =
\left(
\begin{array}{cccc}
m_{\tilde{\gamma}} &        m_{az} &      0 &   0 \\
            m_{az} & m_{\tilde{z}} &  m_{Z} &   0 \\
0 & m_{Z} &   \mu \sin 2\beta & - \mu \cos 2\beta \\
0 &     0 & - \mu \cos 2\beta & - \mu \sin 2\beta
\end{array}
\right)
\enspace ,
\end{equation}
where
\begin{eqnarray}
m_{\tilde{\gamma}}
&=& M_2 \sin^{2}\theta_{W}
 + M_1 \cos^{2}\theta_{W}
\enspace ,
\nonumber \\
m_{\tilde{z}}
&=& M_2 \cos^{2}\theta_{W}
 + M_1 \sin^{2}\theta_{W}
\enspace ,
\\
m_{az}
&=& \sin\theta_{W} \cos\theta_{W} ( M_2 - M_1 )
\enspace .
\nonumber
\end{eqnarray}


\section{Loop Functions}

In this appendix, we collect the different loop functions in the text.

The loop functions for triangle diagrams, entering in EDMs, $b \to s
\gamma$ and the anomalous magnetic moment are,

\begin{eqnarray}
  F_1(x)   &=&
        \frac{1}{12(x-1)^4}(x^3-6x^2+3x+2+6x\ln x),
\\
  F_2(x)   &=&
        \frac{1}{12(x-1)^4}(2x^3+3x^2-6x+1-6x^2\ln x),
\\
  F_3(x)   &=&
        \frac{1}{2(x-1)^3}(x^2-4x+3+2\ln x),
\\
  F_4(x)   &=&
        \frac{1}{2(x-1)^3}(x^2-1-2x\ln x),
\end{eqnarray}

The loop functions for box diagrams, entering in $\varepsilon_K$,
$\Delta M_{B_d}$ and $\Delta M_{B_s}$, are,
\begin{eqnarray}
G(a,b) & = & -\Frac{1}{4} a b \left( {a^2 -8 a + 4\over (a-b)(a-1)^2}\ln{a}
\ + \  {b^2 -8 b + 4\over (b-a)(b-1)^2}\ln{b}\ -\ {3\over (a-1)(b-1)} \right)
\end{eqnarray}

\begin{eqnarray}
Y_1(a,b,c,d)& = & {a^2\over (b-a)(c-a)(d-a)}\ln{a}
    \ +\ {b^2\over (a-b)(d-b)(d-b)}\ln{b}  
                                             \nonumber \\
 & & +{c^2\over (a-c)(b-c)(d-c)}\ln{c}
     \ + \ {d^2\over (a-d)(b-d)(c-d)}\ln{d}
\end{eqnarray}
and
\begin{eqnarray}
Y_2(a,b,c,d) &=&  \sqrt{4 c d}\Bigg[{a\over (b-a)(c-a)(d-a)}\ln {a}
   \  +\ {b\over (a-b)(c-b)(d-b)}\ln{b}
                                              \nonumber \\
 & & +{c\over (a-c)(b-c)(d-c)}\ln{c}
   \  +\ {d\over (a-d)(b-d)(c-d)}\ln{d}\Bigg]\,.
\end{eqnarray}

\begin{table}
\caption{
Approximate solutions of third generation squark--mass parameters and
Higgs mass parameters,
$m^2_i =$ $c_{i1} M^2_{10}$ $ + c_{i2} M^2_5$
       $+ c_{i3} M^2_{H1}$ $+ c_{i4} M^2_{H2}$
       $+ c_{i5} M^2_{1/2}$
       $+ c_{6i} \mbox{Re}(A_d(0) M^*_{1/2})$ $+ c_{i7} \mbox{Re}(A_u(0) M^*_{1/2})$
       $+ c_{8i} |A_d(0)|^2$$ + c_{i9} |A_u(0)|^2$.}
\begin{tabular}{cc|cccc|ccc|cc} 
 & & $c_{i1}$ & $c_{i2}$ & $c_{i3}$ & $c_{i4}$ & $c_{i5}$ &
     $c_{i6}$ & $c_{i7}$ & $c_{i8}$ & $c_{i9}$ \\ \hline
$\tan\beta=2.5$ 
 & $m^2_{D_3}$ & 0    & 1    & 0    & 0    & 6.14
               & 0    & 0    & 0    & 0    \\ 
 & $m^2_{U_3}$ & 0.43 & 0    & 0    &-0.28 & 3.94
               & 0    & 0.18 & 0    &-0.04 \\
 & $m^2_{Q_3}$ & 0.72 & 0    & 0    &-0.14 & 5.49
               & 0    & 0.09 & 0    &-0.02 \\
 & $m^2_{H_1}$ & 0    & 0    & 1    & 0    & 0.36
               & 0    & 0    & 0    & 0    \\ 
 & $m^2_{H_2}$ &-0.85 & 0    & 0    & 0.58 &-3.05
               & 0    & 0.28 & 0    &-0.06 \\ \hline 
$\tan\beta=5$ 
 & $m^2_{D_3}$ & 0    & 1    & 0    & 0    & 5.83
               & 0.01 & 0    & 0    & 0    \\ 
 & $m^2_{U_3}$ & 0.49 & 0    & 0    &-0.25 & 3.88
               & 0    & 0.25 & 0    &-0.06 \\
 & $m^2_{Q_3}$ & 0.74 & 0    & 0    &-0.13 & 5.30
               & 0    & 0.12 & 0    &-0.03 \\
 & $m^2_{H_1}$ & 0    & 0    & 1    & 0    & 0.29
               & 0.01 & 0    & 0    & 0    \\ 
 & $m^2_{H_2}$ &-0.76 & 0    & 0    & 0.62 &-2.76
               & 0    & 0.37 & 0    &-0.09 \\ \hline 
$\tan\beta=10$ 
 & $m^2_{D_3}$ &-0.01 & 0.99 &-0.01 & 0    & 5.77
               & 0.03 & 0    &-0.01 & 0    \\ 
 & $m^2_{U_3}$ & 0.50 & 0    & 0    &-0.25 & 3.91
               & 0    & 0.26 & 0    &-0.06 \\
 & $m^2_{Q_3}$ & 0.75 & 0    & 0    &-0.12 & 5.29
               & 0.01 & 0.13 & 0    &-0.03 \\
 & $m^2_{H_1}$ &-0.01 &-0.01 & 0.99 & 0    & 0.21
               & 0.04 & 0    &-0.01 & 0    \\ 
 & $m^2_{H_2}$ &-0.75 & 0    & 0    & 0.63 &-2.71
               & 0    & 0.39 & 0    &-0.09 \\ \hline 
$\tan\beta=30$ 
 & $m^2_{D_3}$ &-0.08 & 0.92 &-0.08 & 0    & 5.12
               & 0.23 & 0    &-0.06 & 0    \\ 
 & $m^2_{U_3}$ & 0.50 & 0    & 0    &-0.25 & 3.93
               & 0    & 0.26 & 0    &-0.06 \\
 & $m^2_{Q_3}$ & 0.71 &-0.04 &-0.04 &-0.12 & 4.97
               & 0.12 & 0.13 &-0.03 &-0.03 \\
 & $m^2_{H_1}$ &-0.14 &-0.14 & 0.86 & 0    &-0.77
               & 0.37 & 0    &-0.11 & 0    \\ 
 & $m^2_{H_2}$ &-0.75 & 0    & 0    & 0.63 &-2.66
               & 0    & 0.39 & 0    &-0.09 \\ \hline 
$\tan\beta=40$ 
 & $m^2_{D_3}$ &-0.16 & 0.84 &-0.16 & 0    & 4.46
               & 0.32 & 0    &-0.08 & 0    \\ 
 & $m^2_{U_3}$ & 0.49 & 0    & 0    &-0.25 & 3.92
               & 0    & 0.25 & 0    &-0.06 \\
 & $m^2_{Q_3}$ & 0.66 &-0.08 &-0.08 &-0.13 & 4.63
               & 0.16 & 0.12 &-0.04 &-0.03 \\
 & $m^2_{H_1}$ &-0.30 &-0.30 & 0.70 & 0    &-1.79
               & 0.52 & 0    &-0.17 & 0    \\ 
 & $m^2_{H_2}$ &-0.76 & 0    & 0    & 0.62 &-2.66
               & 0    & 0.37 & 0    &-0.09 
\end{tabular}
\label{tab:ScalarMass1}
\end{table}
\begin{table}
\caption{
Approximate solutions of $|\mu|^2$ and $m_{A^0}$ in the CMSSM case:
$m^2_i =  c_{i1} M^2_{0}$ $ +  c_{i2} M^2_{1/2}
       + c_{3i} A_0 M_{1/2}+ c_{i4} A_0^2 + c_{i5}$. 
$c_{i5}=-m^2_Z/2$ in case of $|\mu|^2$ and $m^2_W-m^2_Z$ in case of 
$m^2_{H^+}$.}
\begin{tabular}{cc|cccc} 
$\tan \beta$ & & $c_1$ & $c_2$ & $c_3$ & $c_4$ \\ \hline
2.5 & $|\mu|^2$ & 0.51 & 3.70 & -0.33 & 0.08  \\ 
  & $m^2_{H^+}$ & 1.75 & 4.71 & -0.38 & 0.09  \\ \hline
  5 & $|\mu|^2$ & 0.19 & 2.89 & -0.39 & 0.09  \\ 
  & $m^2_{H^+}$ & 1.23 & 3.30 & -0.39 & 0.09  \\ \hline
 10 & $|\mu|^2$ & 0.13 & 2.74 & -0.40 & 0.10  \\ 
  & $m^2_{H^+}$ & 1.10 & 2.98 & -0.36 & 0.08  \\ \hline
 30 & $|\mu|^2$ & 0.12 & 2.66 & -0.39 & 0.09  \\ 
  & $m^2_{H^+}$ & 0.69 & 1.89 & -0.02 &-0.02  \\ \hline
 40 & $|\mu|^2$ & 0.15 & 2.66 & -0.37 & 0.09  \\ 
  & $m^2_{H^+}$ & 0.24 & 0.87 &  0.15 &-0.08  \\ \hline
\end{tabular}
\label{tab:ScalarMass3}
\end{table}
\begin{table}
\caption{
Approximate solutions of $|\mu|^2$ and $m_{A^0}$ in the $SU(5)$ case:
$m^2_i =  c_{i1} M^2_{10}$ $ + c_{i2} M^2_5
       + c_{i3} M^2_{H1} + c_{i4} M^2_{H2}
       + c_{i5} M^2_{1/2}
       + c_{6i} A_d(0) M_{1/2}+ c_{i7} A_u(0) M_{1/2}
       + c_{8i} A^2_d(0) + c_{i9} A^2_u(0)  + c_{i10}$.
$c_{i10}=-m^2_Z/2$ in case of $|\mu|^2$ and $m^2_W-m^2_Z$ in case of 
$m^2_{H^+}$.}
\begin{tabular}{cc|cccc|ccc|cc} 
$\tan \beta$ & & $c_1$ & $c_2$ & $c_3$ & $c_4$ & $c_5$
               & $c_6$ & $c_7$ & $c_8$ & $c_9$  \\ \hline
2.5 & $|\mu|^2$ & 1.01 & 0    & 0.19 &-0.69 & 3.70
                & 0    &-0.33 & 0    & 0.08 \\ 
  & $m^2_{H^+}$ & 1.17 & 0    & 1.38 &-0.80 & 4.71
                & 0    &-0.38 & 0    & 0.09 \\ \hline
  5 & $|\mu|^2$ & 0.80 & 0    & 0.04 &-0.64 & 2.89
                & 0    &-0.39 & 0    & 0.09  \\ 
  & $m^2_{H^+}$ & 0.82 & 0    & 1.08 &-0.67 & 3.30
                & 0.01 &-0.40 & 0    & 0.10  \\ \hline
 10 & $|\mu|^2$ & 0.75 & 0    & 0.01 &-0.63 & 2.74
                & 0    &-0.40 & 0    & 0.10 \\ 
  & $m^2_{H^+}$ & 0.75 &-0.01 & 1.01 &-0.64 & 2.98
                & 0.04 &-0.40 &-0.01 & 0.10  \\ \hline
 30 & $|\mu|^2$ & 0.75 & 0    & 0    &-0.63 & 2.66
                & 0    &-0.39 & 0    & 0.09  \\ 
  & $m^2_{H^+}$ & 0.60 &-0.14 & 0.85 &-0.63 & 1.89
                & 0.37 &-0.39 &-0.11 & 0.09  \\ \hline
 40 & $|\mu|^2$ & 0.76 & 0    & 0    &-0.62 & 2.66
                & 0    &-0.37 & 0    & 0.09  \\ 
  & $m^2_{H^+}$ & 0.46 &-0.30 & 0.70 &-0.62 & 0.87
                & 0.52 &-0.37 &-0.17 & 0.09 
\end{tabular}
\label{tab:ScalarMass2}
\end{table}
\begin{table}
\caption{
Approximate solutions of first generation squark--mass parameters,
$m^2_i =  b_{i1} M^2_{10} + b_{i2} M^2_5 + b_{i3} M^2_{1/2}$ }
\begin{tabular}{cccc} 
  & $b_{i1}$  & $b_{i2}$  & $b_{i3}$ \\ \hline
$m^2_{D_1}$  & 0 & 1 & 6.1  \\ 
$m^2_{U_1}$  & 1 & 0 & 6.15 \\
$m^2_{Q_1}$  & 1 & 0 & 6.5  \\
$m^2_{E_1}$  & 1 & 0 & 0.15 \\
$m^2_{L_1}$  & 0 & 1 & 1.5 \\
\end{tabular}
\label{tab:sfermionfirst}
\end{table}
\begin{table}
\caption{Approximate solutions of the A--parameters as a function of
the GUT--parameters for various $\tan\beta$. All Parameters are given by
$A_i = a_{i1} A_d(0) + a_{i2} A_u(0) + a_{i3}M_{1/2}$ }
\begin{tabular}{c|ccc||ccc||ccc||ccc||ccc}
     & \multicolumn{3}{c||}{$\tan\beta=2.5$} &
       \multicolumn{3}{c||}{$\tan\beta=5$}  &
       \multicolumn{3}{c||}{$\tan\beta=10$}  &
       \multicolumn{3}{c||}{$\tan\beta=30$}  &
        \multicolumn{3}{c}{$\tan\beta=40$} \\
 & $a_{i1}$ & $a_{i2}$ & $a_{i3}$  & $a_{i1}$ & $a_{i2}$ & $a_{i3}$
 & $a_{i1}$ & $a_{i2}$ & $a_{i3}$  & $a_{i1}$ & $a_{i2}$ & $a_{i3}$
 & $a_{i1}$ & $a_{i2}$ & $a_{i3}$   \\ \hline
 $A_u$ & 0 & 0.58 & -2.89 & 0    & 0.63 & -2.88 & 0    & 0.63 & -2.90 & 
      0    & 0.63 & -2.92 & 0    & 0.62 & -2.91 \\
 $A_d$ & 1 & 0    & -3.74 & 1    & 0    & -3.63 & 0.99 & 0    & -3.61 &
      0.86 & 0    & -3.42 & 0.7  & 0    & -3.15  \\
 $A_t$ & 0 & 0.15 & -1.98 & 0    & 0.24 & -2.09 & 0    & 0.25 & -2.12 &
     -0.04 & 0.25 & -2.07 & -0.08 & 0.24 & -1.97  \\
 $A_b$ & 1 &-0.14 & -3.45 & 0.99 &-0.13 & -3.36 & 0.98 &-0.12 & -3.33 &
      0.74 &-0.12 & -2.93 & 0.45 &-0.13 & -2.41
\end{tabular}
\label{tab:Aparam}
\end{table}

\end{document}